\documentclass[times,twocolumn,final]{elsarticle}

\usepackage{multirow}
\usepackage{multicol}
\usepackage{amsmath}
\usepackage{mathtools}

\usepackage{setspace}

\usepackage{medima}
\usepackage{framed}

\usepackage[T1]{fontenc}
\usepackage[utf8]{inputenc}

\usepackage{graphicx}
\usepackage[export]{adjustbox}

\usepackage{ragged2e}

\usepackage{amssymb}
\usepackage{pifont}
\usepackage{amssymb}
\usepackage{latexsym}
\usepackage{xurl}
\usepackage[T1]{fontenc}
\usepackage[utf8]{inputenc}

\usepackage{array}
\newcolumntype{H}{>{\setbox0=\hbox\bgroup}c<{\egroup}@{}}

\usepackage{url}
\usepackage{xcolor}

\usepackage{algorithm}
\usepackage{algpseudocode}

\definecolor{newcolor}{rgb}{.8,.349,.1}

\journal{Author Version (preprint)}

\begin{document}

\verso{Bill Cassidy \textit{et~al.}}

\begin{frontmatter}

\title{An Enhanced Harmonic Densely Connected Hybrid Transformer Network Architecture for Chronic Wound Segmentation Utilising Multi-Colour Space Tensor Merging}

\author[1]{Bill \snm{Cassidy}}


\author[1]{Christian \snm{Mcbride}}
\author[1]{Connah \snm{Kendrick}}
\author[2]{Neil D. \snm{Reeves}}
\author[3]{Joseph M. \snm{Pappachan}}
\author[4]{Cornelius J. \snm{Fernandez}}
\author[5]{Elias \snm{Chacko}}
\author[6,7,8]{Raphael \snm{Br\"ungel}}
\author[6,7]{Christoph M. \snm{Friedrich}}
\author[9]{Metib \snm{Alotaibi}}
\author[9]{Abdullah Abdulaziz \snm{AlWabel}}
\author[9]{Mohammad \snm{Alderwish}}
\author[10]{Kuan-Ying \snm{Lai}}
\author[1,3]{Moi Hoon \snm{Yap}\corref{cor1}}
\cortext[cor1]{Corresponding author:}
\ead{m.yap@mmu.ac.uk}

\address[1]{Department of Computing and Mathematics, Manchester Metropolitan University, Dalton Building, Chester Street, Manchester, M1 5GD, UK}
\address[2]{Medical School, Faculty of Health and Medicine, Health Innovation Campus, Lancaster University, LA1 4YW, UK}
\address[3]{Lancashire Teaching Hospitals NHS Foundation Trust, Preston, PR2 9HT, UK}
\address[4]{United Lincolnshire Hospitals NHS Trust, Greetwell Road, Lincoln, LN2 5QY, UK}
\address[5]{Jersey General Hospital, St Helier, JE1 3QS, Jersey}

\address[6]{Department of Computer Science, University of Applied Sciences and Arts Dortmund (FH Dortmund), Emil-Figge-Str. 42, 44227 Dortmund, Germany}
\address[7]{Institute for Medical Informatics, Biometry and Epidemiology (IMIBE), University Hospital Essen, Zweigertstr. 37, 45130 Essen, Germany}
\address[8] {Institute for Artificial Intelligence in Medicine (IKIM), University Hospital Essen, Girardetstr. 2, 45131 Essen, Germany}
\address[9] {University Diabetes Center, King Saud University Medical City, Riyadh, Saudi Arabia}
\address[10]{Independent researcher, Taiwan}

\received{XX XXX 2024}
\finalform{XX XXX 20XX}
\accepted{XX XXX 20XX}
\availableonline{XX XXX 20XX}

\begin{abstract}
Chronic wounds and associated complications present ever growing burdens for clinics and hospitals world wide. 
Venous, arterial, diabetic, and pressure wounds are becoming increasingly common globally. These conditions can result in highly debilitating repercussions for those affected, with limb amputations and increased mortality risk resulting from infection becoming more common. New methods to assist clinicians in chronic wound care are therefore vital to maintain high quality care standards. 
This paper presents an improved HarDNet segmentation architecture which integrates a contrast-eliminating component in the initial layers of the network to enhance feature learning. 
We also utilise a multi-colour space tensor merging process and adjust the harmonic shape of the convolution blocks to facilitate these additional features. 
We train our proposed model using wound images from light-skinned patients and test the model on two test sets (one set with ground truth, and one without) comprising only darker-skinned cases. Subjective ratings are obtained from clinical wound experts with intraclass correlation coefficient used to determine inter-rater reliability. 
For the dark-skin tone test set with ground truth, we demonstrate improvements in terms of Dice similarity coefficient ($+0.1221$) and intersection over union ($+0.1274$). 
Measures from the qualitative analysis also indicate improvements in terms of high expert ratings, with improvements of $>3$\% demonstrated when comparing the baseline model with the proposed model. 
This paper presents the first study to focus on darker-skin tones for chronic wound segmentation using models trained only on wound images exhibiting lighter skin. Diabetes is highly prevalent in countries where patients have darker skin tones, highlighting the need for a greater focus on such cases. Additionally, we conduct the largest qualitative study to date for chronic wound segmentation. 
All source code for this study is available at: https://github.com/mmu-dermatology-research/hardnet-cws 





\end{abstract}


\end{frontmatter}


\section{Introduction}
\label{sec1}
Diabetes is now regarded as a global epidemic, resulting in most part from a systematic increase in populations becoming overweight and obese (\cite{moura2019dfu}). Programmes that target the condition have historically shown only short-term benefits, with longer-term effects yet to be established (\cite{khuntie2012diabetes, davies2017diabetes}). The situation is similar for obesity (\cite{ong2023diabetes}), a common factor in diabetes occurrence (\cite{klein2022obesity}). Arterial leg ulcers (ALUs) and diabetic foot ulcers (DFUs) are a debilitating and costly complication of diabetes (\cite{moura2019dfu}), with recent findings suggesting an association between DFU episodes and all-cause resource utilisation and increased mortality risk (\cite{petersen2022mortality}). Venous leg ulcers (VLUs) and pressure ulcers (PRUs) are the most common types of complex skin ulcers (\cite{jenkins2019ulcers}), with ulcer prevalence in the diabetic population estimated to be 13\% in North America (\cite{zhang2017ulceration}). The incidence of chronic wounds is high and is estimated to continue on an upward trajectory (\cite{eriksson2022wounds}).

Patients diagnosed with DFU are two to three times more likely to die than patients without and are predisposed to numerous comorbidities, including peripheral artery disease, cardiovascular disease, neuropathy, retinopathy, and nephropathy. VLUs and DFUs often result in significantly impaired quality of life (\cite{franks2016venous, mader2019dfu, xiong2020diabetes}). Occurrence of ulcers is linked to an increased incidence of both amputation and mortality, especially in the presence of advanced age, peripheral artery disease and anemia (\cite{franks2016venous, costa2017mortality, vainieri2020mortality}). Chronic wounds exert a significant physical and emotional burden on patients (\cite{renner2017depression, polikandrioti2020depression}), with depression being associated with an increased risk at initial and subsequent occurrence (\cite{iversen2015depression, iversen2020telemedicine}).

Chronic wounds are typically correlated with comorbidities such as diabetes, vascular deficits, hypertension, and chronic kidney disease (\cite{sen2021wound}). Diabetic neuropathy is highly prevalent in DFU cases and is the primary cause of DFU formation (\cite{petrone2021foot}), meaning that patients have lost sensation in their foot due to nerve damage (\cite{rathur2007neuropathic}). This means that patients often go through long periods not realising they have a DFU until the wound becomes much worse and leads to other serious complications. Infection affects more than 50\% of all DFU cases (\cite{bader2008diabetic}) and represents one of the most common causes of diabetes related hospitalisation (\cite{petrone2021foot}). Diabetic leg and foot ulcers are amongst the most expensive wound types to treat in the United States (\cite{sen2021wound}). For VLUs, the recurrence rate within 3 months after wound closure is as high as 70\% (\cite{franks2016venous}). 

Management of chronic wounds can be a long and difficult task, for both patient and clinician. This is especially true for wounds that are not caught early, and require more intensive treatment programmes. This can mean frequent visits to clinics or hospitals for assessment by experts (\cite{boulton2005burden, netten2017mobile}). Even after accomplished wound healing, recurrences are frequent and often lead to minor or major amputation of lower extremities (\cite{apelqvist1993prognosis, larsson1998prognosis}). The post COVID-19 climate poses further risks and challenges to the treatment of chronic wounds, given that diabetic patients are placed in the high-risk category. To this end, recent years have seen an increased research interest in the remote detection and monitoring of wounds using non-contact methods (\cite{cassidy2021cloudbased, reeves2021diabetes, joseph2022future}). 

Evolving current telemedicine systems to include remote wound monitoring represents an opportunity to reduce risks to vulnerable patients and to ease significantly overburdened healthcare systems (\cite{yammine2021telemedicine}). Furthermore, the advent of cheap consumer mobile devices and easily accessible cloud platforms promotes the idea of making these technologies available to poorer regions, where patients may experience reduced access to expert healthcare providers. 
Low cost, easy-to-use non-invasive devices that can detect and monitor wounds could act as a mechanism to promote patient engagement with the monitoring of their health. 

A growing body of evidence has shown the ability of convolutional neural networks (CNNs) to equal or surpass experienced dermatologists for detection and classification in related domains (\cite{esteva2017dermatologist, brinker2019classification, brinker2019superior, fujisawa2019surpasses, pham2020binary, jinnai2020pigmented, haenssle2021classification}). In this regard, deep learning may be able to assist in providing more objective results in domains which are prone to high levels of subjectivity. Changes to wound area have been shown to be a robust predictor in healing status (\cite{sheehan2003healing}). Segmentation of chronic wounds allows for more accurate assessment of changes to wound shape and size over time when compared to more generalised localisation techniques. 
In the next section, we discuss the recent notable developments in this domain.

\section{Related Work}
\label{sec2}
Studies on deep learning tasks related to chronic wounds have become a growing interest in the research community in recent years due to the possible benefits that such technologies might offer in real-world clinical settings (\cite{goyal2018robust, cassidy2023eval}). In this section, we examine the more prominent studies conducted in chronic wound segmentation research that have helped to guide the experiments presented in this paper. 

\cite{goyal2017fully} were one of the first research groups to investigate chronic wound segmentation using convolutional neural networks (CNNs). They trained a number of fully convolutional networks (FCN) to segment DFU wounds and associated periwounds using a dataset comprising 600 DFU images together with ground truth masks which were provided by wound experts at Lancashire Teaching Hospitals (LTH), UK. A two-tier transfer learning approach using two publicly available general image datasets was used - Pascal VOC and ImageNet segmentation datasets. The DFU segmentation dataset was divided into 420 training images, 60 validation images, 120 test images, and 105 images of healthy feet. In the joint segmentation of wound and periwound regions the highest performing model was FCN32-s with a Dice similarity coefficient (DSC) of 0.899. For segmentation of ulcer regions only, the highest performing model was FCN-16s, reporting a DSC of 0.794. For segmentation of only periwounds, the highest performing model was FCN-16s, reporting a DSC of 0.851. This work noted that the FCN-AlexNet and FCN-32s models were less accurate in the segmentation of irregular boundaries, and that the smaller pixel strides used in FCN-16s and FCN-8s resulted in improved detection of such examples. This study also observed an overlap of periwound and wound regions in prediction results due to ambiguities in feature boundaries. A limitation of this work is the small number of samples used in the experiments, which may make the results difficult to generalise across more diverse datasets. 
\cite{wang2020segmentation} conducted wound segmentation experiments using MobileNetV2, which was pretrained using the Pascal VOC segmentation dataset. For training and testing, they used a newly introduced dataset of 1109 DFU images ($train = 831; test = 278$). A localisation method was used as a preprocessing stage to exclude non-DFU wound regions from images before the segmentation stage. As a post-processing step, morphological algorithms were used (small region removal and hole-filling). Their test results reported a mean DSC of 0.9047. However, this work presents several limitations. First, all wound images were very small patches that are heavily padded to a resolution of $224 \times 224$ pixels. Wound pixels therefore comprised only very small regions of the images. Excluding padding, the average size of the wound regions in the training set is $71 \times 104$ pixels, and the average wound region size in the test set is $70 \times 101$ pixels. 
At such low resolutions, as small as $17 \times 18$ pixels, a large number of wound features may be lost. 
They also tested their model on the Medetec dataset, and obtained a DSC of 0.9405. 

In later works, \cite{wang2022fuseg} conducted the Foot Ulcer Segmentation Challenge (FUSC) 2021 whereby a new DFU dataset was released ($train = 810$, $val = 200$, $test = 200$). 
This new dataset comprised of examples with less significant padding compared to their prior dataset, with images exhibiting more foot and background features. The winner of the FUSC 2021, \cite{mahbod2021automatic}, achieved an image-based DSC of 0.8880, which was 1.67\% lower than the prior DSC reported by \cite{wang2020segmentation}. This may indicate that the task was more difficult when larger wound images were introduced. In the FUSC 2021, models were required to learn features that are more complex that were absent from the prior experiments conducted by \cite{wang2022fuseg} which used a smaller dataset comprising notably smaller wound regions and thus fewer features. 
\cite{scebba2021segmentation} noted the numerous challenges associated with wound segmentation, including wound type heterogeneity, variance in tissue colouration, wound shapes, background features, anatomical location, variety of image capturing scenarios, and non-standard specifications of capture devices. They observed that standardisation initiatives in medical wound photography may lead to additional workload burdens on clinical routine, and that the proposal of standards would likely not result in a desired consistent approach in real-world scenarios. Their proposed method utilised a MobileNet localisation model to assist a U-Net segmentation model to reduce non-wound features. 
This study used a total of five chronic wound datasets (1) SwissWOU - a private dataset of DFU ($n = 1096$) and systemic sclerosis digital ulcers ($n = 63$), (2) SIH (second healing intention dataset) ($n = 58$) (\cite{yang2016wound}), (3) DFUC2020 ($n = 2000$) (\cite{cassidy2021eval}), (4) FUSC ($n = 60$) (\cite{wang2022fuseg}), (5) Medetec ($n = 53$) (\cite{thomas2014medetec}). We observe that for some of the datasets used in this study, complete sets were not utilised in the experiments. For the FUSC, Medetec, and SIH datasets, only a selection of images were used. The authors experimented using a range of well-known segmentation networks, both with and without localisation preprocessing (manual and automated). 
When tested using only the SwissWOU DFU images (10\% of all patients), their results showed that U-Net was the highest performing network ($MCC = 0.85$, $IoU = 0.75$). Their test results for the SwissWOU systemic digital ulcers, Medetec, SIH, and FUSC images also showed U-Net to be the best performing network ($MCC = 0.8725$, $IoU = 0.7875$). 
HarDNet-DFUS (Harmonic Densely Connected Network), proposed by \cite{liao2022seg}, was the winning entry for the DFUC2022, achieving a DSC of 0.7287. The design is based on a prior work, HarDNet-MSEG (\cite{huang2021mseg}), and is the basis of our proposed methods in the present paper. HarDNet-DFUS uses inter-layer connections which were configured according to the required block depth $n$. Therefore, when $n = 9$, the resulting factors are 1, 3, and 9, allowing for shortcuts to the 1st, 3rd, and 9th convolutions. This results in the removal of the power of 2 constraint found in the original block design. A block depth of 3, 9, and 15 was selected for the final design, replacing the original depth of 4, 9, and 16. This results in reduced data movement using the same number of convolutional layers. Additionally, they replaced the receptive field blocks (RFB) in the decoder with a large window attention (Lawin) transformer. The original HarDNet network mainly utilised $3\times3$ convolutions to increase computational density, which changes the model from being memory-bound to compute-bound (\cite{chao2019hardnet}). To increase accuracy further, they used an ensembling strategy using 5-fold cross validation and test time augmentation (TTA). Augmented images were added to the test set when testing the sub-models, with the output averages used as the final prediction results. However, they found that this method was not consistent, and would sometimes degrade performance in terms of DSC and IoU.

\cite{ramachandram2022wound} proposed a chronic wound segmentation network for tissue type segmentation (AutoTissue) and wound segmentation (AutoTrace) designed for use in a commercial mobile app. The AutoTrace model implemented a typical auto-encoder design using depth-wise separable convolution layers, attention gates, and strided depth-wise convolutions resulting in downsample activations which act as an alternative to fixed max-pooling. Additive attention gates were added to skip connections to regulate activations from previous network layers. 
Bilinear upsampling was used in the decoder blocks followed by depth-wise separable convolution layers, helping to reduce memory requirements. The AutoTissue segmentation model implemented EfficientNetB0 as the encoder path, with a decoder comprising 4 layers with each layer utilising two-dimensional bilinear upsampling followed by 2 depth-wise convolution layers. AutoTrace was trained with a private dataset comprising 467,000 wound images, while AutoTissue was trained with a second private dataset comprising 17,000 wound images. For both datasets, both images and ground truth labels were obtained from hospitals in North America, allowing for a diverse range of wound images. However, details were not disclosed regarding the exact composition of the datasets. The study reported an mIoU of 0.8644 for wound segmentation and an mIoU of 0.7192 for tissue and wound segmentation. Clinicians rated 91\% (53/58) of the results as between fair and good for segmentation and tissue segmentation quality. Qualitative assessment of is rare chronic wound related deep learning studies. However, the sample size used is limited, whereby only 58 examples were rated. 


\cite{swerdlow2023pressure} used a private dataset exhibiting stages 1-4 PRUs, acquired from eKare Inc. Mask R-CNN with a ResNet101 backbone was trained for segmentation and classification of each PRU stage of development. 
The dataset comprised 969 PRU images ($train = 848$, $test = 121$).  
The study reported a DSC of 0.92 for stage 1 PRU, 0.85 for stage 2 PRU, 0.93 for stage 3 PRU, and 0.91 for stage 4 PRU. 
The wound image acquisition protocol indicated that images be taken from approximately 40-65 cm distance from the wound. Additionally, the study excluded PRU wounds that were smaller than $2 \times 2$ cm, which may have limited testing of the model's true ability to segment a range of wound sizes. 

The use of different colour spaces in CNNs was explored by \cite{gowda2019colornet}. Their classification experiments on the CIFAR-10, CIFAR-100, SVHN, and ImageNet datasets showed that different classes were sensitive to models trained on different colour spaces. They trained a series of DenseNet models using multiple image datasets that had been converted to different colour spaces, with each DenseNet using a different colour space as input. The outputs from each DenseNet were then used as input into a final dense layer to generate weighted predictions from each sub-DenseNet. Increased computational overhead, a result of using multiple DenseNets, was addressed by using smaller and wider DenseNets. This work showed that training with images from multiple colour spaces provided comparable results to significantly larger models, such as DenseNet-BC-190-40, with a reduction of more than 10M parameters. 


In later CNN-based colour space studies, \cite{simon2022deeplumina} trained classification models using RGB and luminance images. Their experiments utilised a ResNet101 pretrained model for feature learning and an SVM for the classifier. They trained and tested their model with the Describable Texture Dataset (DTD) and the Flickr Material Dataset (FMD). Compared to prior works, for the DTD, they reported an accuracy improvement of 0.73\%, and for the FMD they reported an accuracy improvement of 6.95\%.

In more recent work, \cite{mcbride2024colour} conducted preliminary experiments which merged individual colour channels from different colour spaces into single tensors when training a chronic wound U-Net segmentation model. They found that different colour channel merging operations using RGB, CIELAB, and YCrCb colour spaces improved segmentation performance by 0.0264 for IoU and 0.0348 for DSC when testing on the FUSC dataset. However, this study was limited by the use of only a simple U-Net model. 

One of the most prominent aspects of chronic wound research in deep learning, as highlighted by our literature review, has been a lack of substantial publicly available fully annotated datasets. Another notable factor in the field is a lack of focus on patients exhibiting darker skin tones. The biases towards lighter skin tones present in deep learning models in dermatology research is well established (\cite{wen2021datasets}). \cite{bencevic2024bias} observed significant bias in skin lesion segmentation against darker-skin cases when performing in and out-of-sample evaluation. 
Furthermore, they also found that methods used to mitigate bias do not result in significant bias reduction. Most of the publicly available chronic wound datasets comprise cases that were collected from lighter skin patients. While some datasets do contain examples with darker skin tones, these are not quantified. In the next section, we discuss the chronic wound datasets that we used in our experiments.

\section{Chronic Wound Datasets}
\label{sec3}
Large medical imaging datasets present notable challenges when used to train deep learning networks (\cite{wen2021datasets}). Issues such as image duplication, image and feature similarity (\cite{dipto2023similarity}), varying image quality, label noise and the presence of visual artefacts can significantly impact model performance (\cite{akkoca2020objects, cassidy2021isic, daneshjou2021guidelines, winkler2021scalebars, jaworek2023skinhair, pewton2024dca}). 


Our research group has been responsible for the release of the first substantial publicly available DFU wound datasets with ground truth labels (\cite{cassidy2020dfuc, yap2021classification, kendrick2022segmentation}). With the release of each dataset, we have conducted yearly challenges in association with the International Conference on Medical Image Computing and Computer Assisted Intervention (\cite{cassidy2020dfuc, yap2021evaluation, cassidy2021eval, yap2022overview, yap2024report}). Our datasets comprise of over 20,000 high quality DFU wound photographs together internationally coordinated clinical labelling provided by experts in podiatry. 
Table \ref{table:datasets} shows a summary of all the datasets used in our chronic wound segmentation experiments. We use 10 public datasets, 1 private dataset, and a dataset comprising Google Image Search images which we collected using the Creative Commons License search option to remove copyrighted images from search results. These images vary significantly, both in size and quality. To obtain these images, we used search terms such as ``diabetic foot ulcer", ``neuropathic ulcer", ``venous ulcer", ``pressure ulcer", ``wound", and ``chronic wound".

The private dataset used in our experiments is the The King Saud University Medical City (KSUMC) dataset. This dataset comprises 115 DFU wound images and was obtained from the King Saud University Medical City, Saudi Arabia. 
The images were acquired using a Fujifilm Finepix SL260 digital camera at various resolutions and orientations. The KSUMC dataset was obtained with ethical approval from King Saud University Medical City, Saudi Arabia (REF: 24/1159/IRB). 


\begin{table*}
    \centering
    \caption{A summary of public and private wound image datasets used in our experiments. Note that the Train, Val, and Test columns show how the datasets were originally divided. YWHD - Yang Wound Healing dataset; AZH - Advancing the Zenith of Healthcare Wound Care dataset; FUSC - Foot Ulcer Segmentation Challenge dataset; GIS-W - Google Image Search wound images; CWDB - Complex Wound DB; Wseg - Wound Segmentation dataset; KSUMC - King Saud University Medical City dataset; Cla - classification; Seg - segmentation; Mul - multimodal.}
    \label{table:datasets}
    \scalebox{1.0}{
    \begin{tabular}{|p{3.75cm}|p{2.1cm}|p{2.4cm}|p{0.79cm}|p{0.79cm}|p{0.69cm}|p{1.1cm}|p{1.1cm}|p{1.1cm}|}
		\hline
		Publication                    & Name     & Resolution                    & Task             & Train & Val & Test & Total & Status \\ \hline\hline
		\cite{thomas2014medetec}       & Medetec  & $560\times(347-444)$          & Seg              &       & -   & -    & 608   & Public \\ \hline
		\cite{yang2016wound}           & YWHD     & $5184\times3456$              & -                & -     & -   & -    & 201   & Public \\ \hline
		\cite{alzubaidi2020dfu_qutnet} & Alzubaidi & various                      & Cla              & -     & -   & -    & 493   & Public \\ \hline
		\cite{wang2020segmentation}    & AZH      & $224\times224$                & Seg              & 831\textsuperscript{$\sharp$} & - & 278\textsuperscript{$\sharp$} & 1109 & Public \\ \hline
            \cite{kendrick2022segmentation} & DFUC2022\textsuperscript{*} & $640\times480$ & Seg & 2000\textsuperscript{$\sharp$}  & - & 2000\textsuperscript{$\sharp$} & 4000 & Public \\ \hline
		\cite{wang2024fusc}          & FUSC    & $512\times512$                & Seg               & 810\textsuperscript{$\sharp$}   & 200\textsuperscript{$\sharp$} & 200  & 1210 & Public \\ \hline 

            \cite{groh2021evaluating}      & Fitzpatrick17k & various                & -                 & -   & - & -  & 16,529 & Public \\ \hline
  
		\cite{krecichwost2021woundsdb} & WoundsDB & $4896\times3264$              & Mul              & -     & -   & -    & 188\textsuperscript{$\sharp$} & Public  \\ \hline
		- (2023)                       & GIS-W    & various                       & -                & -     & -   & -    & 186 & Public  \\ \hline
            \cite{pereira2022complexwounddb} & CWDB   & various             & Seg              & -     & -   & -    & 27\textsuperscript{$\sharp$} & Public \\ \hline
            \cite{oota2023wsnet} & Wseg     & $331\times331$                & Seg              & -     & -   & -    & 2686 & Public \\ \hline
            - (2024) & KSUMC     & various                & Mul              & -     & -   & -    & 115 & Private \\ \hline
		
		\multicolumn{9}{@{}l}{* includes pathology class and anatomical location labels. \textsuperscript{$\sharp$} includes ground truth masks available to the present study.}
	\end{tabular}
	}
\end{table*}

\subsection{Expert Wound Delineation}
All training, validation and test cases for the DFUC2022 dataset were delineated with the location of DFUs in polygon coordinates. 
The VGG Image Annotator tool (\cite{dutta2016via, dutta2019vgg}) was used to delineate images with polygons indicating the ulcer region. The ground truth was produced by five healthcare professionals who specialise in treating diabetic foot ulcers and associated pathology, comprising consultant physicians and podiatrists, all with more than 5 years professional experience. The instruction for annotation was to delineate each DFU with a polygon region. 


We evaluate the agreement between the expert annotators on 800 cases (20\% of the data) chosen at random using the Jaccard Similarity Index (JSI) and DSC. The DSC of the delineation between experts is 0.6981$\pm$0.2544, the JSI is 0.5876$\pm$0.2670, and accuracy is 0.9869$\pm$0.0291. 

\begin{figure}[!h]
	\centering
	\begin{tabular}{ccccc}
		\includegraphics[width=1.3cm,height=1.6cm]{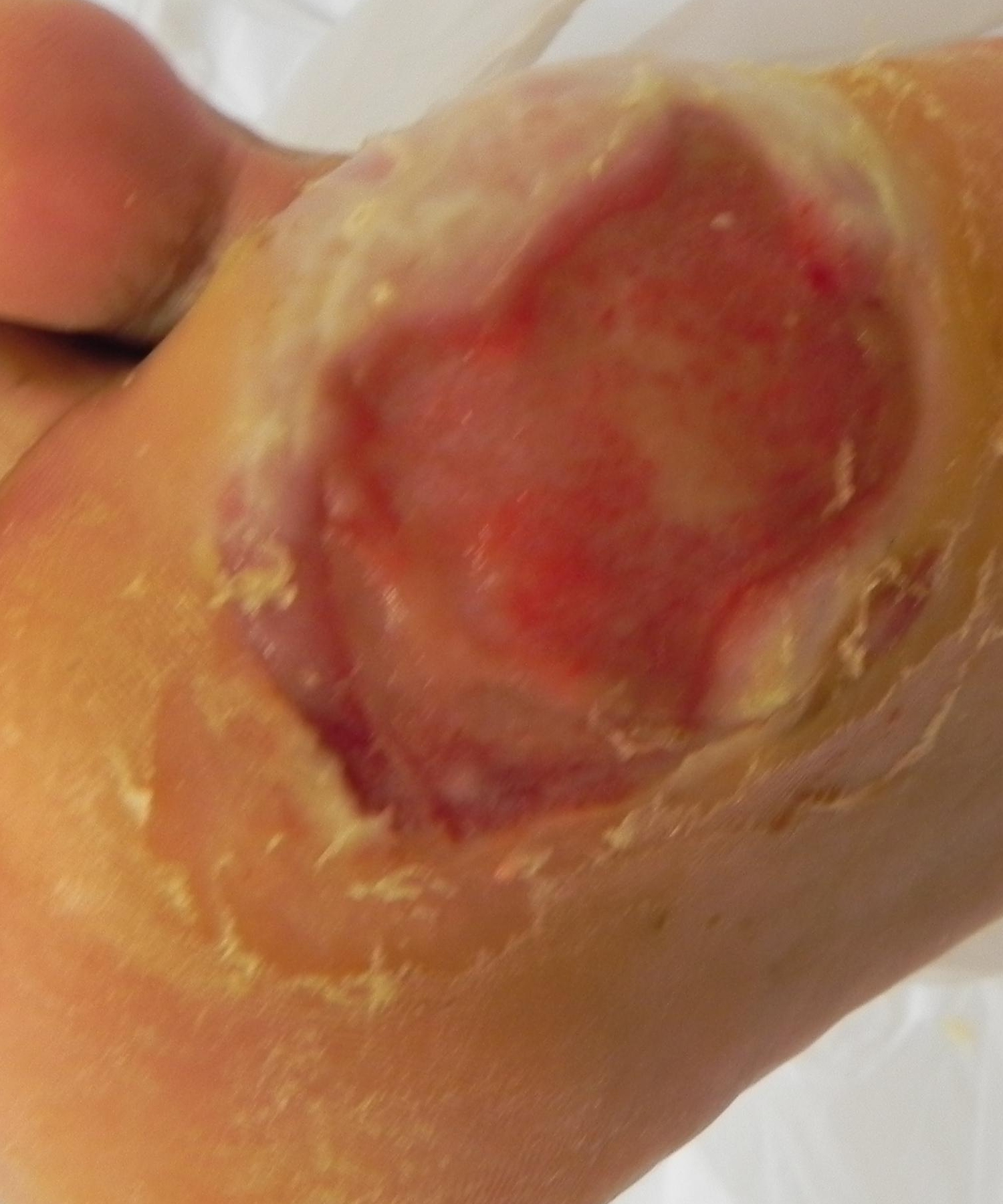} &
		\includegraphics[width=1.3cm,height=1.6cm]{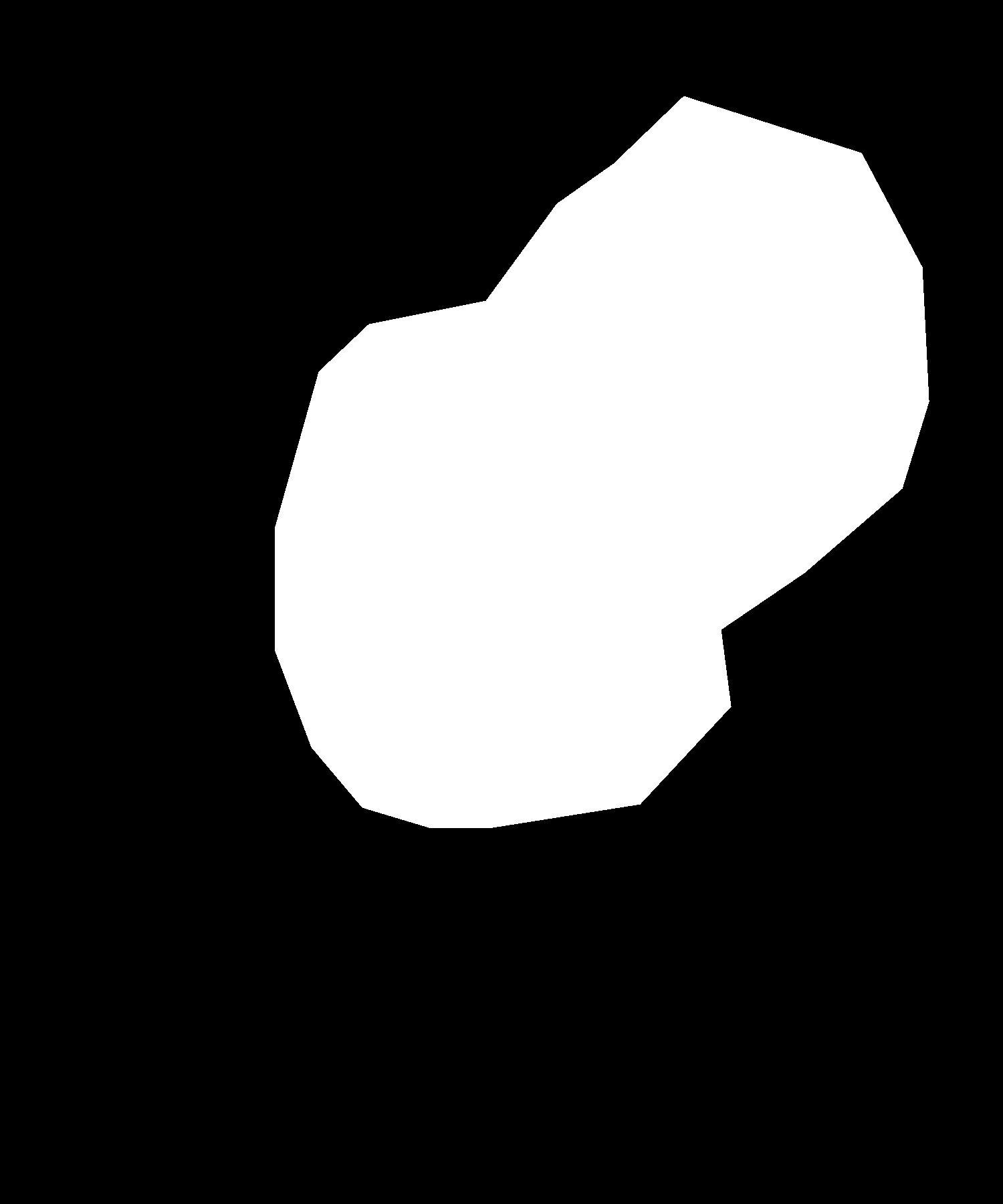} &
		\includegraphics[width=1.3cm,height=1.6cm]{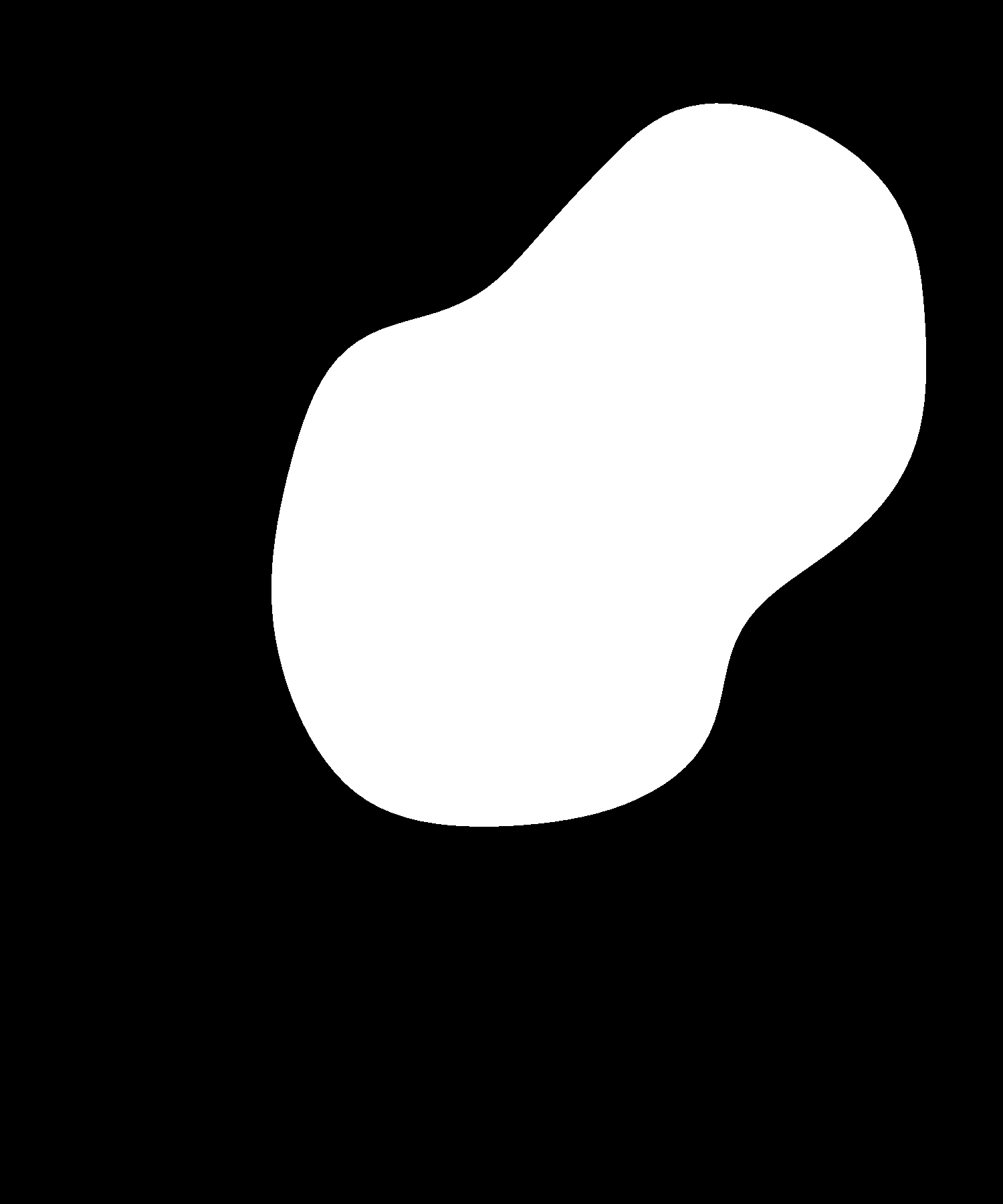} &
		\includegraphics[width=1.3cm,height=1.6cm]{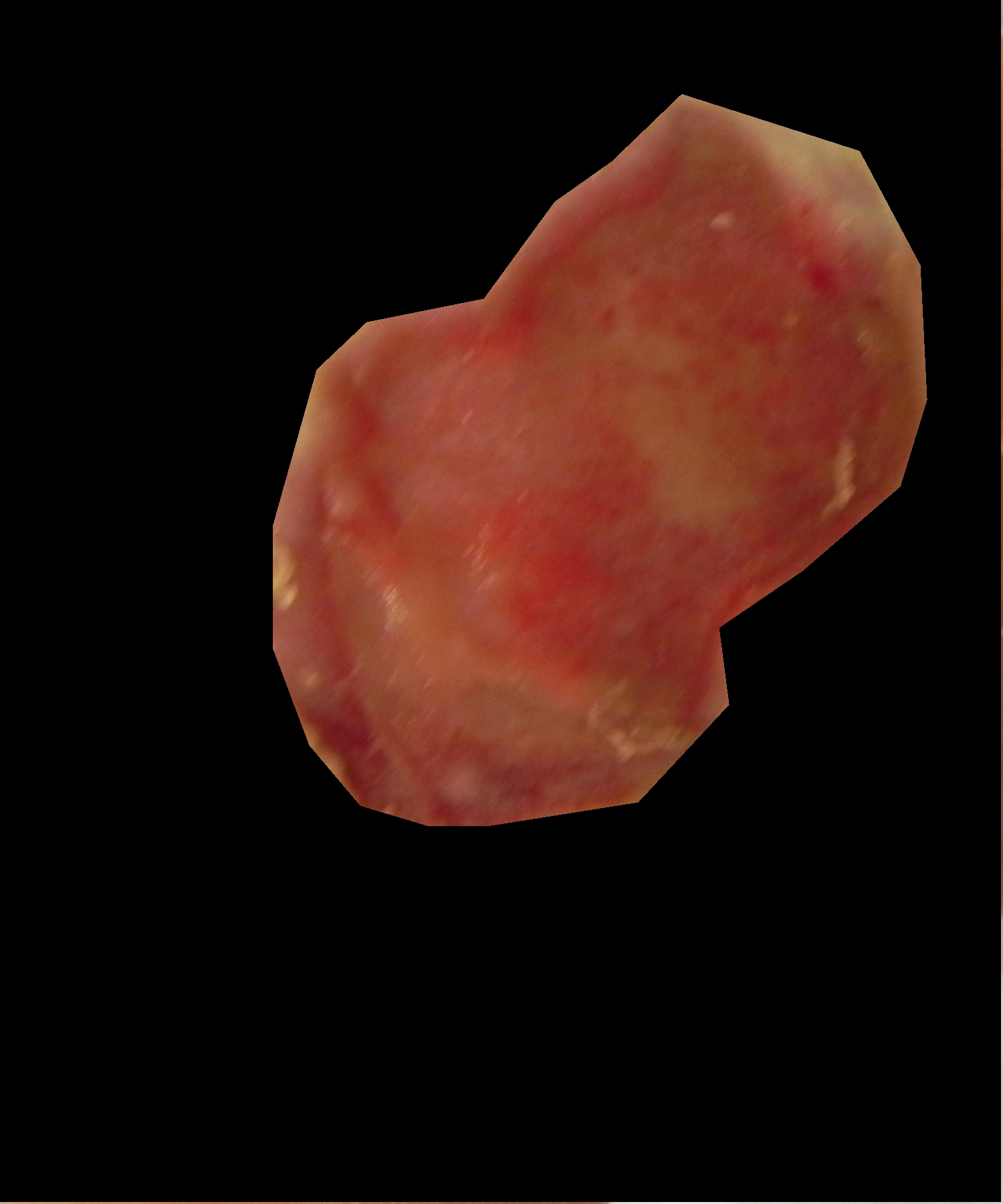} &
		\includegraphics[width=1.3cm,height=1.6cm]{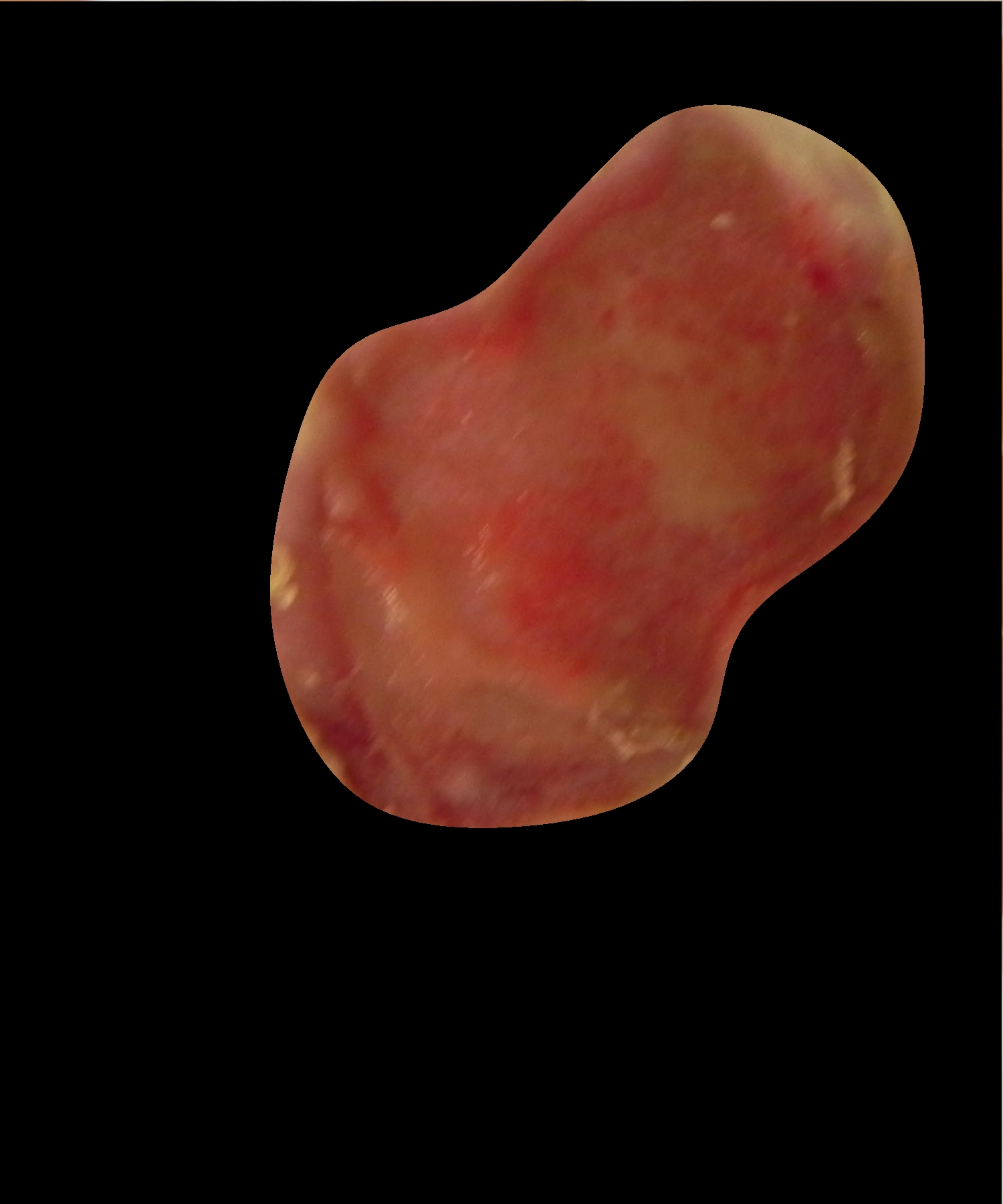} \\
		(a) & (b) & (c) & (d) & (e) \\  
	\end{tabular}
	\caption[]{Illustration of an image from the DFUC2022 training set and corresponding masks: (a) original image; (b) original mask based on clinician delineation; (c) original mask processed using active contour model; (d) original image with clinician delineation mask overlaid; and (e) original image with original mask processed using active contour model overlaid. Note that images were cropped for illustration purposes.}
	\label{fig:masks}
\end{figure}

The use of active contour masks when used as ground truth has been shown to provide superior agreement with machine predicted results in chronic wound segmentation tasks (\cite{kendrick2022segmentation}). Therefore, in our experiments, for the DFUC2022 dataset we use ground truth masks that have been processing using the original polygon delineations with an active contour model applied to smooth delineated vertices. The active contour model masks were produced using the MATLAB (The MathWorks, Inc., Massachusetts) method created by \cite{kroon2022snake}, using default parameters. Figure \ref{fig:masks} shows an example of the two different mask types applied to a training image from the DFUC2022 dataset. To further validate that the smoothing effect did not alter the delineation of the experts, we measure the similarity of the masks produced by the clinicians and the masks post-processed by active contour on the training set. The DSC is 0.9620$\pm$0.0259, the JSI is 0.9279$\pm$0.0462, and the accuracy is 0.9991$\pm$0.0012. These evaluations support our statement that the pre-processing stage has provided a smoothing effect, but did not alter the experts' delineation. 

\section{Method}
This section details the training, validation, and testing workflow, proposed model architecture, and corresponding metrics used for our segmentation experiments. 

\subsection{Metrics}
We utilised a series of widely used evaluation metrics to determine the accuracy of the models trained, validated, and tested in our wound segmentation experiments. Intersection over union (IoU) and DSC were selected as the main metrics for determining segmentation model accuracy. DSC was chosen for its representation as the harmonic mean of precision and recall, giving a balanced evaluation between false positive and false negative predictions. The relevant mathematical expressions for IoU and DSC are as follows:
\begin{align}
  \begin{split}
    IoU &= \frac{|X \cap  Y|}{|X| \cup |Y|}
    \label{eq:iou}
  \end{split}
  \intertext{}
  \begin{split}
    DSC &= 2 * \frac{|X \cap Y|}{|X| + |Y|}
    \label{eq:jaccard}
  \end{split}
\end{align}

\noindent
where $X$ and $Y$ represent the ground truth mask and predicted mask respectively. 

We also utilise two additional statistical hypothesis testing metrics to better understand the Type I and Type II errors associated with deep learning segmentation algorithm performance. The two additional metrics we use are False Positive Error (FPE) and False Negative Error (FNE) which are defined as follows:
\begin{align}
  \begin{split}
    FPE &= \frac{FP}{FP + TN}
    \label{eq:fpe}
  \end{split}
  \intertext{}
  \begin{split}
    FNE &= \frac{FN}{FN + TP}
   \label{eq:fne}
  \end{split}
\end{align}


\noindent
where $FP$ is the total number of false positive predictions, $TN$ is the total number of true negative predictions, and $FN$ is the total number of false negative predictions.

\subsection{Baseline Experiments}
The first stage in our experiments was to determine the effectiveness of a range of deep learning segmentation networks using the largest publicly available chronic wound dataset (DFUC2022). 
We obtained a series of baselines for training, validation, and test results for a selection of newer segmentation architectures using the DFUC2022 dataset. 
We focus on a selection of more advanced CNN architectures that were not included in the previous baseline experiments reported for DFUC2022 (\cite{yap2024report}). For all baseline experiments, the DFUC2022 dataset images and masks were unchanged from their original resolution ($640\times480$ pixels). A total of 200 images were taken at random from the training set for use as the validation set during training. No augmentation or post-processing methods were used in any of the baseline experiments. All baseline models were trained for 300 epochs with a batch size of 2 using the Adam optimiser with a learning rate of 0.001, 
and a weight decay of 0.0001. All models were trained without the use of pretrained weights. 
The best model for each experiment was selected from the 300 epochs training schedule determined by the highest validation IoU and DSC values. 
The hardware and software configuration for all experiments completed in the present paper was as follows: Debian GNU/Linux 10 (buster) operating system, AMD Ryzen 9 3900X 12-Core CPU, 128GB RAM, NVIDIA GeForce RTX 3090 24GB GPU. Models were trained with Tensorflow 2.4.1 and Pytorch 1.13.1 using Python 3.7.13. 

\begin{table*}
    \centering
    \caption{Baseline results for a selection of deep learning segmentation networks trained, validated and tested on the DFUC2022 dataset (image size = $640 \times 480$ pixels). IoU - intersection over union; DSC - Dice similarity coefficient; FPE - false positive error; FNE - false negative error; DCSA - deeper more compact split-attention; MBS - multi-branch segmentation; EffNet - EfficientNet. ConvNeXt U-Net was trained using the \emph{convnext\_base} backbone. Note that none of the networks evaluated used pretraining.}
    \label{table:baselines}
    \scalebox{0.75}{
    \begin{tabular}{|p{2.9cm}|p{2.4cm}|p{0.85cm}|p{1.4cm}|p{1.5cm}|p{1.5cm}|p{1.1cm}|p{1.2cm}|p{1.3cm}|p{1.2cm}|p{1.4cm}|p{1cm}|p{1cm}|}
        \hline
	Model             & Implementation              & Epoch & Train IoU & Train Loss & Train DSC & Val IoU & Val Loss & Val DSC & Test IoU & Test DSC & FPE     & FNE    \\ \hline \hline
        ResUNet++         & \cite{jha2019resunetpp}     & 152   & 0.6015    & 0.3238     & 0.7213    & 0.4495  & 0.6245   & 0.5767  & 0.3798   & 0.4969   & 0.4315  & 0.3967 \\
        
        U-Net++           & \cite{4uiiurz12020unet}     & 279   & 0.6694    & 0.2662     & 0.7826    & 0.5147  & 0.4505   & 0.6451  & 0.3996   & 0.5179   & 0.4057  & 0.4152 \\
        
        Attention U-Net   & \cite{czekalski2020attunet} & 65    & 0.6710    & 0.2671     & 0.7835    & 0.5352  & 0.4157   & 0.6552  & 0.4135   & 0.5302   & 0.3760  & 0.4238 \\
        
        DCSAU-Net         & \cite{xu2023dcsaunet}       & 245   & 0.5657    & 0.3653     & 0.6887    & 0.4467  & 0.5395   & 0.5712  & 0.3627   & 0.4736   & 0.4498  & 0.4298 \\
        
        MBSNet            & \cite{jin2023seg}           & 81    & 0.6979    & 0.2332     & 0.7999    & 0.5195  & 0.4524   & 0.6446  & 0.3977   & 0.5102   & 0.4240  & 0.3979 \\
        
        ResNet50 U-Net    & \cite{li2023seg}    & 196   & 0.6424    & 0.2892     & 0.7578    & 0.4924  & 0.4612   & 0.6211  & 0.3732   & 0.4915   & 0.3878  & 0.4712 \\
        
        MobileNetV2 U-Net & \cite{li2023seg}    & 34    & 0.6884    & 0.2485     & 0.7946    & 0.5624  & 0.3912   & 0.6844  & 0.4406   & 0.5597   & 0.3542  & 0.3975 \\ 
        
        ConvNeXt U-Net    & \cite{mayali2023bbunet}     & 98    & 0.5529    & 0.3574     & 0.6869    & 0.4339  & 0.5016   & 0.5728  & 0.3087   & 0.4289   & 0.4157  & 0.5476 \\
        
        EffNetB0 U-Net    & \cite{mayali2023bbunet}     & 258   & 0.7817    & \textbf{0.1558}     & 0.8686    & 0.5846  & 0.3693   & 0.7044  & 0.4616   & 0.5784   & 0.3474  & 0.3813 \\
        
        EffNetB1 U-Net    & \cite{mayali2023bbunet}     & 38    & 0.6856    & 0.2388     & 0.7935    & 0.5844  & \textbf{0.3485}   & 0.7038  & 0.4584   & 0.5785   & 0.3396  & 0.3807 \\ 
        
        EffNetB2 U-Net    & \cite{mayali2023bbunet}     & 184   & 0.7575    & 0.1748     & 0.8515    & 0.5843  & 0.3613   & 0.7026  & 0.4461   & 0.5641   & 0.3648  & 0.3828 \\
    
        UNeXt             & \cite{valanarasu2022unext}  & 96    & 0.4580    & 0.4844     & 0.5895    & 0.4398  & 0.5128   & 0.5695  & 0.3383   & 0.4596   & 0.0464  & 0.4660 \\

        HarDNet-DFUS      & \cite{liao2022hardnetdfus}  & 33    & \textbf{0.7889}    & 0.2601     & \textbf{0.8743}    & \textbf{0.6068}  & 0.4610   & \textbf{0.7101}  & \textbf{0.5421}   & \textbf{0.6520}   & \textbf{0.0255}  & \textbf{0.3278} \\

        \hline
	\end{tabular}
	}
\end{table*}

The results of the baseline experiments are summarised in Table \ref{table:baselines}. HarDNet-DFUS is clearly the best overall performing network in terms of training ($IoU = 0.7889, DSC = 0.8743$), validation ($IoU = 0.6068, DSC = 0.7101$), and test metrics ($IoU = 0.5421, DSC = 0.6520, FPE = 0.0255$, $FNE = 0.3278$). 
We observe that the EfficientNet U-Nets record lower training and validation loss rates at 0.1558 (EffNetB0 U-Net) and 0.3485 (EffNetB1 U-Net) respectively. These loss rates are significant, a reduction of 0.1043 for B0 train loss and a reduction of 0.1125 for B1 validation loss. However, these performance gains are not reflected in the test loss results when comparing the EfficientNets with HarDNet. 
The notable differences between validation and test results for the best overall performing network (HarDNet-DFUS) may be indicative of the random nature of the validation set, which might not fully represent the range of features present in the test set. 
We observe that the deeper U-Net variants such as U-Net++ and ResUNet++ demonstrated particularly low metrics, which may be a consequence of the relatively small size of the DFUC2022 dataset and the larger size of these network architectures. 

In addition to the range of network architectures reported on in Table \ref{table:baselines}, we also trained, validated, and tested a number of vision transformer (ViT) segmentation models. However, the test results for the ViTs were well below those reported in Table \ref{table:baselines}. As reported by \cite{zhu2023vit}, ViTs require substantial amounts of training data and are not suitable for use with very small datasets such as those used in the present paper. \cite{zhu2023vit} observed that representation similarity between ViTs trained on small and large datasets comprising of $> 1M$ images differed substantially. They posit that this may be due to a reduction in inductive bias (the relationship between closely positioned input features). Their experiments show that lower layers of ViTs are not able to sufficiently learn local relationships when small amounts of complex data are used. Conversely, recent work by \cite{gani2022vit} suggests that ViTs might be trained on smaller datasets using self-supervised inductive biases. However, even in these scenarios, datasets of up to 100,000 images were used, which although might be considered small in deep learning terms, is still significantly greater than the current publicly available chronic wound datasets. 

\begin{figure}[!h]
  \centering
  \begin{tabular}{ccc}
  (Original Image) & (Ground Truth) & (Prediction) \\
  \includegraphics[width=2.6cm,height=2.2cm]{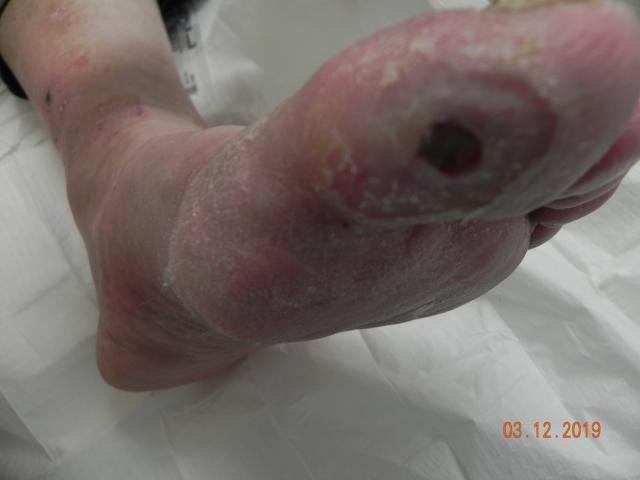} &
  \includegraphics[width=2.6cm,height=2.2cm]{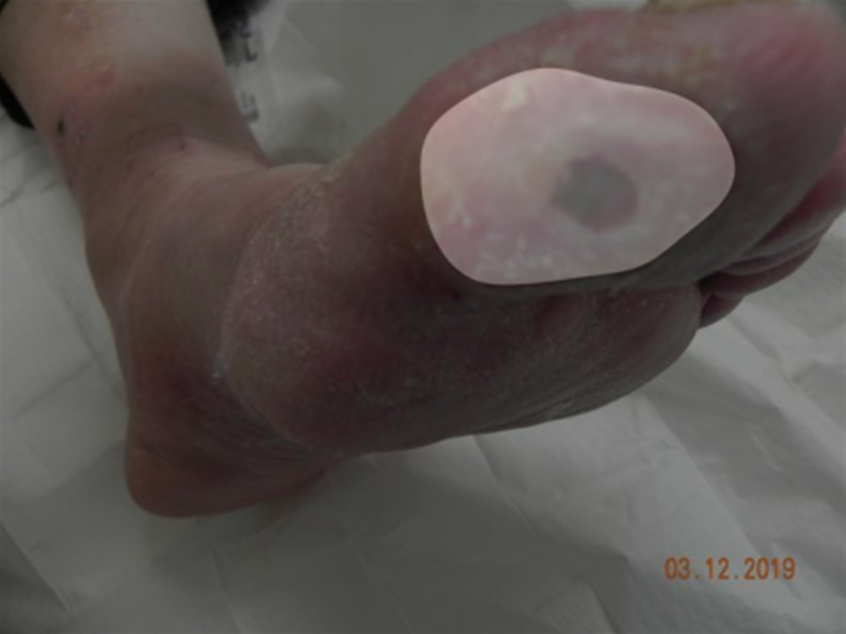} &
  \includegraphics[width=2.6cm,height=2.2cm]{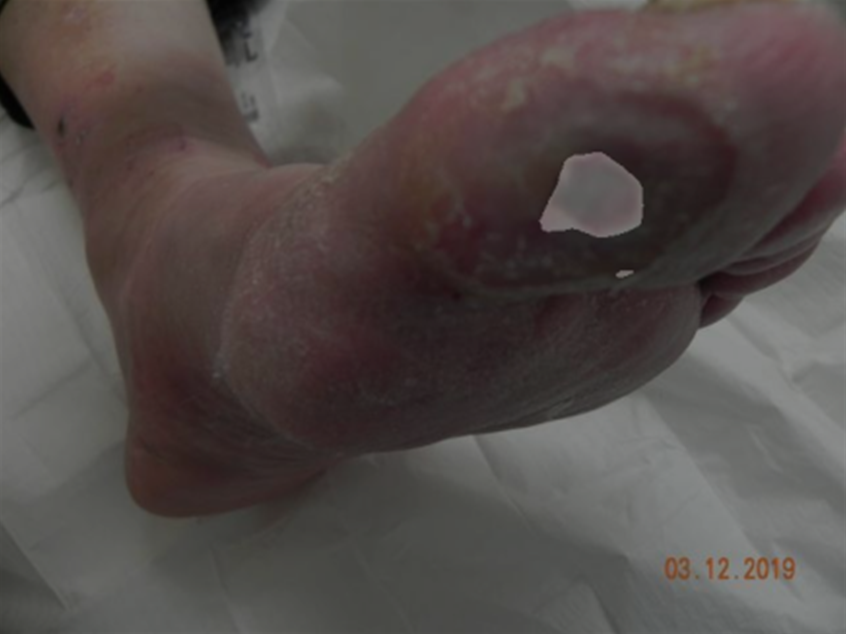} \\

  \includegraphics[width=2.6cm,height=2.2cm]{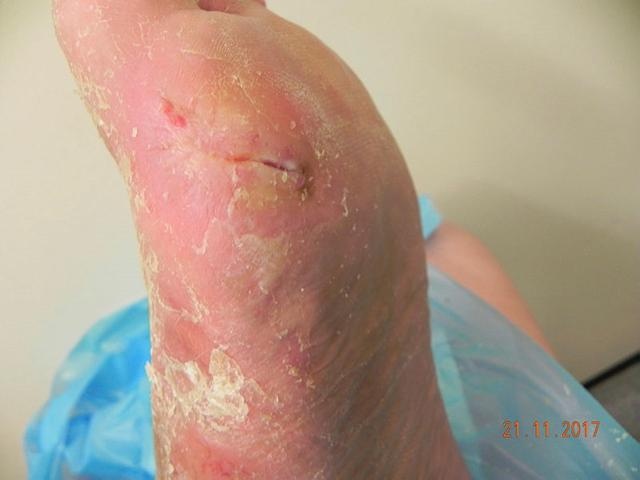} &
  \includegraphics[width=2.6cm,height=2.2cm]{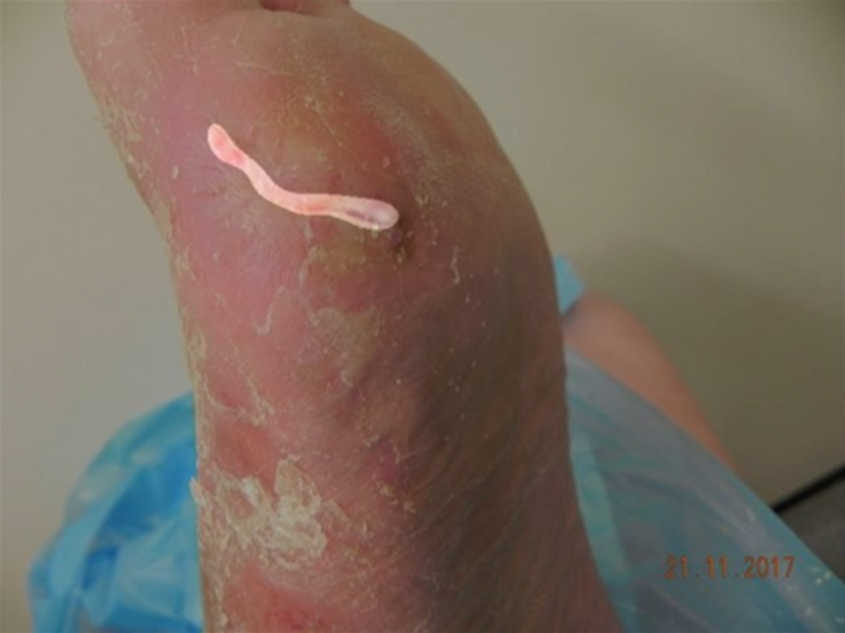} &
  \includegraphics[width=2.6cm,height=2.2cm]{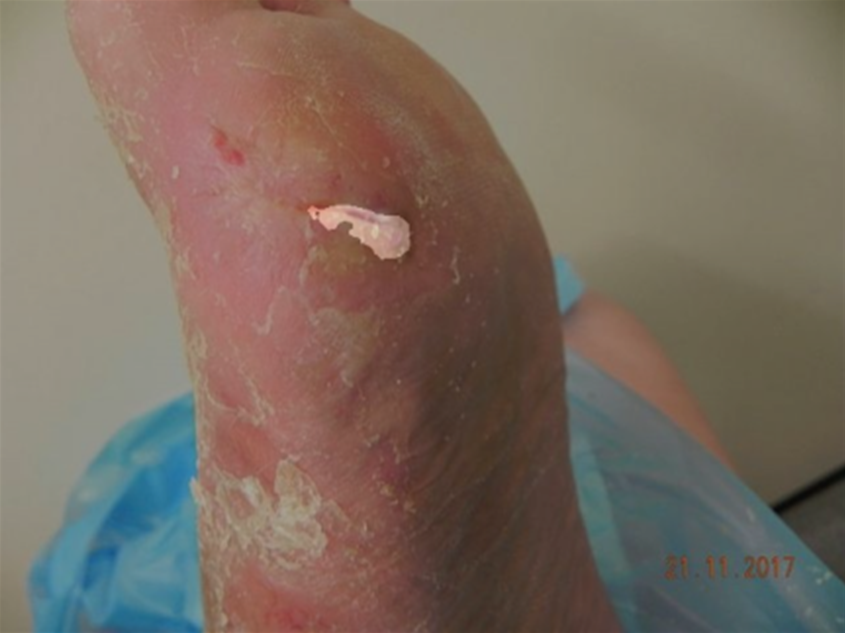} \\

  \includegraphics[width=2.6cm,height=2.2cm]{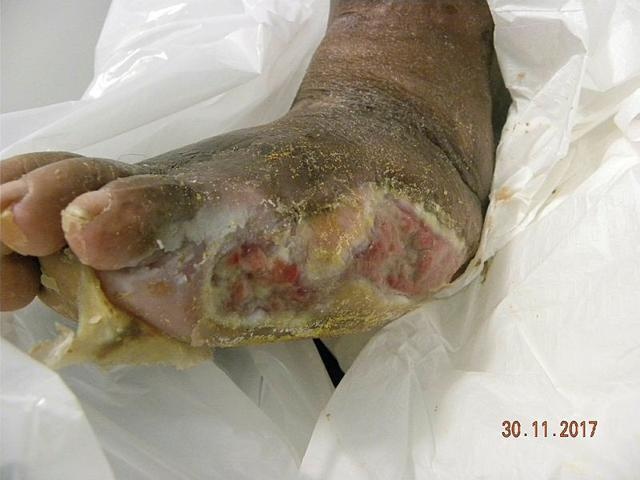} &
  \includegraphics[width=2.6cm,height=2.2cm]{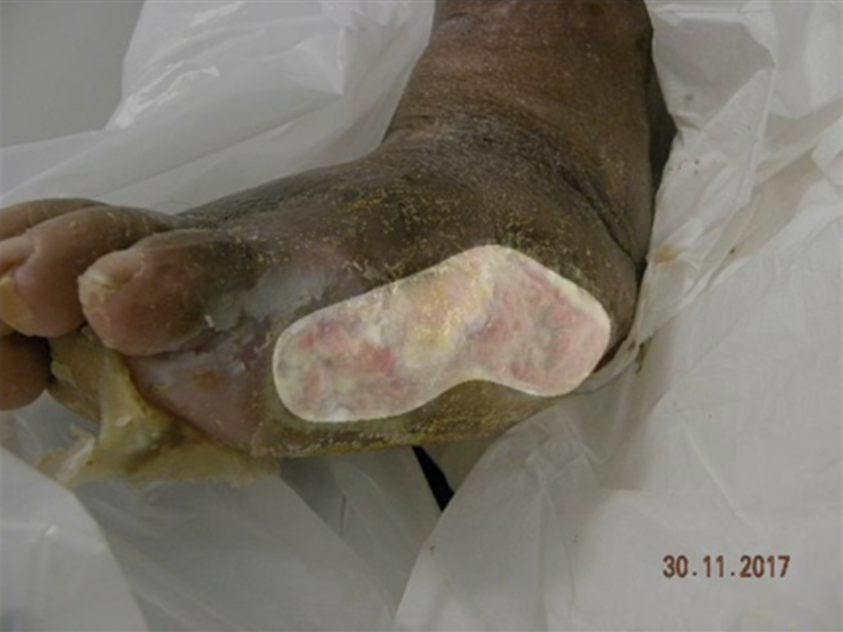} &
  \includegraphics[width=2.6cm,height=2.2cm]{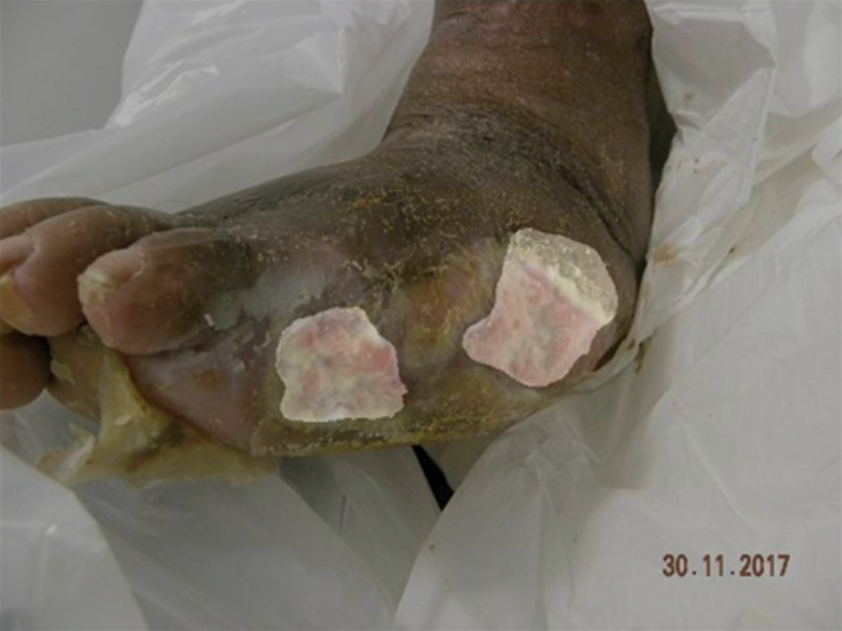} \\
  \end{tabular}
  \caption[]{Illustration of 3 cases from the DFUC2022 dataset where clinical experts determined the baseline model predictions (HarDNet-DFUS) to be superior to the ground truth labels. The first column shows the original images, the second column shows the ground truth, and the third column shows the model predictions.}
  \label{fig:test_variable}
\end{figure}

We compared a selection of ground truth masks with model predictions for the best performing network in the baseline results, which was HarDNet-DFUS. Figure \ref{fig:test_variable} shows 3 cases with original image, ground truth labels, and corresponding baseline model predictions. The first row shows a case where the ground truth mask includes the wound and periwound as a single region, whereas the model predicted only the unhealed wound region. The second row shows a case where the two wound regions are separated by epithelial skin, indicating significant healing between the two non-healed regions. The corresponding prediction shows that only the main wound region was predicted by the model. The third row shows a case where two large wound regions are separated by an epithelial region. The ground truth includes both wound regions and the partially healed region. However, the prediction includes only the non-healed regions. 
These examples demonstrate the significant challenges inherent in human expert wound delineation and how delineation of wound regions can be highly subjective. We asked two clinical experts in wound care (a consultant surgeon and a consultant podiatrist) to indicate agreement with the ground truth labels and corresponding model predictions for the 3 cases shown in Figure \ref{fig:test_variable}. Both experts agreed that the model predictions, although not perfect, were of higher quality than the ground truth labels. Both experts indicated that the automated segmentation of non-healed wound regions was more important than segmentation of healed tissue in terms of automated wound monitoring. We note that these qualitative observations are preliminary and are not to be considered conclusive. The intention is to demonstrate issues present in both expert delineation and limitations of the baseline model. Larger scale qualitative assessment is explored later in the paper. 

We observe that many of the segmentation models that performed highly in other medical imaging domains, such as DCSAU-Net which reported state-of-the-art performance on polyp, multiple myeloma plasma cells, ISIC 2018, and brain tumour segmentation, did not perform well when trained and tested on DFU wounds. We posit that this is due to the larger range of features found across chronic wounds at different stages of development, in addition to the significant visual complexity of such wounds.

\subsection{Construction of Training, Validation, and Test Sets}
The aim of our work is to determine the effectiveness of a segmentation model, trained and validated only on patients with lighter skin tones, to segment wounds on patients with darker skin tones. To this end, we construct a series of datasets for use in our experiments. Our approach for this was to use all publicly available chronic wound datasets that have ground truth masks, together with all datasets that we have access to privately. Wound images were selected based on Fitzpatrick (\cite{fitzpatrick1988skin}) skin types IV (moderate brown skin), V (dark brown skin), and VII (deeply pigmented dark brown or black skin). 
To create the first test set (test set A), we gathered all images with masks exhibiting darker skin tones from the DFUC2022 dataset (68 images and corresponding masks from the training and test sets), the AZH dataset (81 images and corresponding masks from the training and test sets), the CWDB dataset (3 images and corresponding masks), and the FUSC dataset (190 images and corresponding masks from the training and validation sets). Test set A comprises all publicly available wound images with segmentation masks from patients with dark skin tones. 
To create the new training set, we combined the remaining DFUC2022 training and test sets (3893 images and corresponding masks) with 824 images and corresponding masks from the AZH training and test sets. 
For the validation set, we use the remaining 173 AZH images and corresponding masks together with all 24 CWDB images and masks, all 795 FUSC training and validation images and masks, and all 188 WoundsDB images and masks. 
Finally, we created a second test set (test set B) which comprises the same number of images as test set A ($n = 342$) and includes only dark skin tone wound images which do not have ground truth masks which will be assessed qualitatively. Test set B includes wound images from the Alzubaidi dataset ($n = 52$), the Fitzpatrick17k dataset ($n = 4$), the FUSC test set ($n = 35$), the GIS-W dataset ($n = 13$), the Medetec dataset ($n = 8$), the Wseg dataset ($n = 115$), and the KSUMC dataset ($n = 115$). 
A summary of the dataset composition for training, validation, and testing (test set A) is show in Table \ref{table:new_dataset}. 
A summary of test sets A and B is shown in Table \ref{table:test_sets}. 


\begin{table}[!h]
  \centering
  \renewcommand{\arraystretch}{1.0}
  \caption{Summary of the composition of the new dataset used for training, validation, and testing purposes. Note that the training and validation sets comprise only of wound images from light-skinned patients, whereas the test set (test set A) comprises only wound images from patients with darker skin tones.}
  \scalebox{1.0}{
  \label{table:new_dataset}
  \begin{tabular}{|p{2.1cm}|p{1.45cm}|p{1.45cm}|p{1.47cm}|}
    \hline
    Dataset  & Train & Validation & Test Set A \\ \hline \hline 
    DFUC2022 & 3893  & 0          & 68   \\ 
    AZH      & 824   & 173        & 81   \\ 
    CWDB     & 0     & 24         & 3    \\ 
    FUSC     & 0     & 795        & 190  \\
    WoundsDB & 0     & 188        & 0    \\ \hline
    Total    & 4717  & 1180       & 342  \\
    \hline
  \end{tabular}
  }
\end{table}

\begin{table}[!h]
    \centering
    \caption{Summary of the composition of the two dark skin tone test sets used in our experiments. Test set A = 342 images and corresponding masks taken from the DFUC2022, AZH, and FUSC datasets; test set B = 342 images (with no masks) taken from the Alzubaidi, Fitzpatrick17k, FUSC, GIS-W, Medetec, Wseg, and KSUMC datasets.}
    \label{table:test_sets}
    \scalebox{0.85}{
    \begin{tabular}{|p{2.8cm}|p{1.2cm}|p{1.2cm}|p{1.6cm}|}
       \hline
	Dataset Name   & Images & Masks & Test Set \\ \hline\hline
	DFUC2022       & 68     & 68    & A        \\ \hline
	AZH            & 81     & 81    & A        \\ \hline
        CWDB           & 3      & 3     & A        \\ \hline
        FUSC           & 190    & 190   & A        \\ \hline
        
        Alzubaidi      & 52     & 0     & B        \\ \hline
        Fitzpatrick17k & 4      & 0     & B        \\ \hline
        FUSC           & 35     & 0     & B        \\ \hline
        GIS-W          & 13     & 0     & B        \\ \hline
        Medetec        & 8      & 0     & B        \\ \hline
        Wseg           & 115    & 0     & B        \\ \hline
        KSUMC          & 115    & 0     & B        \\ \hline\hline
        Total          & 684    & 342   & A \& B   \\ \hline
	\end{tabular}
	}
\end{table}



\subsection{HarDNet-DFUS Architecture}
Following the analysis of our baseline results, we select the HarDNet-DFUS network architecture used for the winning entry for DFUC2022, proposed by \cite{liao2022hardnetdfus}. This non-symmetrical hybrid transformer segmentation model demonstrated the highest performance in our baseline tests, as shown in Table \ref{table:baselines}, achieving a test DSC of 0.6520 and a test IoU of 0.5421. 
The harmonic element of the network design that is used for the naming of the network is derived from the harmonic pattern of the number of layers used in each HarDNet convolution block. 
In the encoder, HarDNet performs channel splitting on the convolutional outputs in accordance to the number of output connections per layer. This results in an input channel count equal to the number of output channels for each 3x3 convolutional layer. 
The decoder implements a series of Lawin (Large Window Attention) Transformers. Multi-scale features are captured using a Multi-Layer Perception (MLP) decoder and an MLP-Mixer together with Spatial Pyramid Pooling (SPP). 
The MLP-Mixer comprises two layer types: one with MLPs independently applied to image patches for the purpose of mixing per-location features, and a second using MLPs which are applied across patches to enable spatial information to be mixed to enhance spatial representations, as originally proposed by \cite{tolstikhin2021mlpmixer}. 
SPP is a pooling layer with no fixed-size constraints whereby spatial information is retained in local spatial bins where the outputs of each filter are pooled, allowing for multi-scale representations of features (\cite{he2014spp}). 
The decoder design essentially allows for capture of richer contextual data at different scales, utilising transformer elements (Lawin) to focus on improved learning of global relationships. 
Deep supervision is employed in the decoder to aid regularisation in feature learning and to improve convergence behaviour. This involves the use of companion losses which are calculated at different layers in the network, with the final loss calculated as the output loss plus the sum of the companion losses (\cite{lee2015deeply}). Edge loss is also used to enhance the fine-grained details at the edges of prediction masks. 
Finally, an Exponential Moving Average (EMA) function is used during training which maintains moving averages of trainable parameters using an exponential decay. \cite{morales-brotons2024exponential} demonstrated that EMA models generalised better and had improved robustness to noisy labels.

\subsection{HarDNet-CWS Architecture}
We propose a modified HarDNet-DFUS network architecture, henceforth ``HarDNet-CWS'' (Chronic Wound Segmentation), which utilises the following novel enhancements:

\begin{enumerate}
  \item Implementation of an improved multi-colour space tensor merging process that builds on concepts proposed in our previous recent works. 
  \item Modification of the network encoder stem layers using combined instance-batch normalisation in the first encoder block, and switch normalisation in the second encoder block.
  \item Replacement of ReLU6 activation functions with Parameterised Rectified Linear Unit (PReLU) activation functions in all convolution blocks in the encoder. 
  \item Reshaping of the harmonic structure of the HarDNet dense convolution blocks to facilitate the additional colour tensor information.
\end{enumerate}

Each of our proposed enhancements are detailed in the following subsections.


\subsubsection{HarDNet Experimental Setup}
All experiments completed in the following sections used wound images and masks at $640\times480$ pixels. All models were trained for 100 epochs with a batch size of 2 using the AdamW optimiser with a learning rate of 0.00001, 
an epsilon of 0.0000001, and a weight decay of 0.01. The hardware and software configuration used for all experiments is the same as those used for the baseline experiments.

\subsubsection{Multi-colour Space Tensor Merging}
The first adjustment to our proposed HarDNet-CWS model architecture facilitates the range of additional features found in different non-RGB colour spaces. Traditionally, deep learning models that use colour medical photographs are trained and tested using images in the RGB colour space. However, recent preliminary research conducted by \cite{mcbride2024colour} demonstrated that combining individual colour channels from various colour spaces resulted in improved model performance on a range of chronic wound segmentation test sets. Their highest improvement was demonstrated when using the FUSC validation set as an exclusive test set, achieving increases in IoU ($+0.0264$) and DSC ($+0.0348$) when merging RGB colour channels with the Y (luminance) channel from the YCrCb colour space to form a new merged multi-channel tensor (RGB+Y). This work demonstrated that merging individual channels from non-RGB colour spaces resulted in higher performance gains when compared to merging whole colour spaces together. However, a limitation of this work is that it was only demonstrated using a simple U-Net architecture (\cite{ronneberger2015unet}). In this work, we experiment with the colour space channels that demonstrated the highest performance improvements in the prior studies completed by \cite{mcbride2024colour}. We complete experiments that utilise the merging of different colour channel tensors from the RGB, YCbCr, and CIELAB colour spaces. Based on the prior results from the experiments conducted by \cite{mcbride2024colour}, we experiment by merging RGB with the Y luminance channel from the YCbCr colour space, and the `A' chromaticity channel from the CIELAB colour space. We also propose an alternative representation of luminance, which we refer to as exaggerated luminance (eY), which is derived from the RGB colour space. 

For the experiments which focus on merging RGB with the Y and A channels, a summary of results is shown in Table \ref{table:tensor_merging_ay}. Algorithm \ref{alg:tensor_merge_ya} shows the process of merging the RGB channels with the Y and A channels to form newly merged tensors. In terms of test results, the RGB+A, RGB+Y, and RGB+Y+A experiments all show improvements over the baseline results, with the RGB+Y+A experiment demonstrating the highest test set performance increases for test IoU ($+0.0180$), test DSC ($+0.0241$), and FNE ($-0.0055$). 

\begin{table*}[!h]
  \centering
  \caption{Summary of results for the multi-colour channel tensor merging operations when merging RGB colour channels with `A' chromaticity (from the CIELAB colour space) and luminance (Y channel from the YCbCr colour space).}
  \label{table:tensor_merging_ay}
  \scalebox{0.77}{
  \begin{tabular}{|p{3.4cm}|p{1.7cm}|p{1.4cm}|p{1.5cm}|p{1.5cm}|p{1.2cm}|p{1.2cm}|p{1.3cm}|p{1.3cm}|p{1.4cm}|p{1.0cm}|p{1.0cm}|}
  \hline
  Colour Channels & Best Epoch & Train IoU & Train Loss & Train DSC & Val IoU & Val Loss & Val DSC & Test IoU & Test DSC & FPE    & FNE    \\ \hline \hline
  RGB (baseline)  & 50         & \textbf{0.9427}    & \textbf{0.0975}     & \textbf{0.9694}    & 0.6258  & 0.4063   & 0.7176  & 0.5350   & 0.6389   & 0.0597 & 0.3254 \\
  RGB+A           & 19         & 0.7315    & 0.3299     & 0.8340    & 0.6167  & 0.3942   & 0.7066  & 0.5393   & 0.6449   & \textbf{0.0421} & 0.3267 \\
  RGB+Y           & 28         & 0.8380    & 0.2301     & 0.9084    & 0.6140  & 0.3777   & 0.7060  & 0.5402   & 0.6515   & 0.0446 & 0.3580 \\
  RGB+Y+A         & 33         & 0.8745    & 0.1895     & 0.9310    & \textbf{0.6319}  & \textbf{0.3581}   & \textbf{0.7229}  & \textbf{0.5530}   & \textbf{0.6630}   & 0.0508 & \textbf{0.3199}  \\
  \hline
  \end{tabular}
  }
\end{table*}

\begin{algorithm}[!h]
  \caption{RGB+Y+A tensor merging algorithm.}
  \label{alg:tensor_merge_ya}
  \begin{algorithmic}[1]
    \Procedure{Tensor\_Merge}{$rgb\_image$}
    \State $rgb\_tensor \gets to\_tensor(rgb\_image)$
    \State $lab \gets convert\_rgb\_to\_lab(rgb\_tensor)$
    \State $a \gets split(lab)[1]$
    \State $ycrcb \gets convert\_rgb\_to\_ycrcb(rgb\_tensor)$
    \State $y \gets split(ycrcb)[0]$
    \State $image \gets merge([rgb\_tensor, y, a])$
    \State Return $image$
    \EndProcedure
  \end{algorithmic}
\end{algorithm}



To build on the prior tensor merging work completed by \cite{mcbride2024colour}, we experiment further with the Y channel in the tensor merging operation. Our approach was to increase the difference between lighter and darker values in the Y channel by first normalising then applying a fixed exponential. We also experimented by switching the R and B coefficients during the conversion process. The process of deriving the eY channel from the RGB colour space and merging the corresponding tensors is described in Algorithm \ref{alg:tensor_merge_ey}. For all experiments which utilise eY, we use the derivation of luminance equation (see Equation \ref{eq:deriv_y}) as defined by the BT.709-4 standard as proposed by the \cite{itu2000bt}. 

\begin{equation}
Y = 0.2126 R + 0.7152 G + 0.0722 B
\label{eq:deriv_y}
\end{equation}
\noindent
where $R$ represents the red channel value, $G$ represents the green channel value, and $B$ represents the blue channel value. 

\begin{algorithm}[!h]
  \caption{RGB+eY tensor merging algorithm.}
  \label{alg:tensor_merge_ey}
  \begin{algorithmic}[1]
    \Procedure{Tensor\_Merge}{$rgb\_image$}
    \State $rgb\_tensor \gets to\_tensor(rgb\_image)$
    \State $r,g,b \gets split(rgb\_tensor)$
    \State $l \gets (r\times0.0722 + g\times0.7152 + b\times0.2126)$ 
    \State $l \gets to\_array(((l) \div max(l)) \times 255)$
    \State $ey \gets to\_array((l\,^5 \div max(l\,^5)) \times 255)$
    \State $image \gets merge([rgb\_tensor, ey])$
    \State Return $image$
    \EndProcedure
  \end{algorithmic}
\end{algorithm}

Table \ref{table:tensor_merging_ey_coef} shows the results of the eY experiments, with the results compared to the baseline RGB results. The test results for the experiments with and without R \& B coefficient swapping show a clear improvement over both the baseline test results and the RGB+A, RGB+Y, and RGB+Y+A results shown in Table \ref{table:tensor_merging_ay}. Compared to the best results from the prior experiments (see Table \ref{table:tensor_merging_ay}) the RGB+eY tensor merging operation with switched R and B coefficients demonstrate test set performance improvements in terms of test IoU ($+0.0174$), test DSC ($+0.0139$), FPE ($-0.0026$), and FNE ($-0.0461$). 

\begin{table*}[!h]
  \centering
  \caption{Summary of results for the multi-colour channel tensor merging operations when merging RGB colour channels with exaggerated luminance (eY) and `A' chromaticity using normal RGB coefficients (NC) and switched R and B coefficients (SC). Note that when deriving eY from RGB an exponent value of 5 was used for these experiments.}
  \label{table:tensor_merging_ey_coef}
  \scalebox{0.77}{
  \begin{tabular}{|p{3.4cm}|p{1.7cm}|p{1.4cm}|p{1.5cm}|p{1.5cm}|p{1.2cm}|p{1.2cm}|p{1.3cm}|p{1.3cm}|p{1.4cm}|p{1.0cm}|p{1.0cm}|}
  \hline
  Colour Channels & Best Epoch & Train IoU & Train Loss & Train DSC & Val IoU & Val Loss & Val DSC & Test IoU & Test DSC & FPE    & FNE    \\ \hline \hline
  RGB (baseline)  & 50         & \textbf{0.9427}    & \textbf{0.0975}     & \textbf{0.9694}    & 0.6258  & 0.4063   & 0.7176  & 0.5350   & 0.6389   & 0.0597 & 0.3254 \\
  RGB+eY (NC)     & 32         & 0.8670    & 0.1996     & 0.9267    & 0.6224  & 0.3944   & 0.7108  & 0.5422   & 0.6518   & 0.0481 & 0.3245 \\
  RGB+eY (SC)     & 32         & 0.8585    & 0.2089     & 0.9213    & 0.6232  & 0.3903   & 0.7128  & \textbf{0.5576}    & \textbf{0.6654}   & \textbf{0.0420} & 0.3119 \\
  RGB+eY+A (NC)   & 28         & 0.7825    & 0.2877     & 0.8719    & 0.6193  & 0.4033   & 0.7094  & 0.5436   & 0.6484   & 0.0452 & \textbf{0.3108} \\
  RGB+eY+A (SC)   & 26         & 0.8171    & 0.2554     & 0.8957    & \textbf{0.6302}  & \textbf{0.3580}   & \textbf{0.7201}  & 0.5464  & 0.6544    & 0.0423 & 0.3427 \\
  \hline
  \end{tabular}
  }
\end{table*}

\begin{figure*}[!h]
  \centering
  \begin{tabular}{cccc}
  Y & eY & eYS-R\&B & Difference\\
  \includegraphics[width=4cm,height=3.2cm]{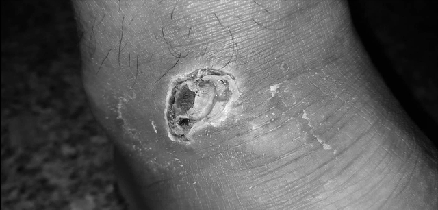} &
  \includegraphics[width=4cm,height=3.2cm]{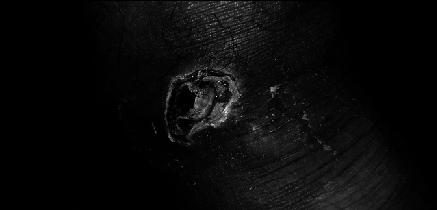} &
  \includegraphics[width=4cm,height=3.2cm]{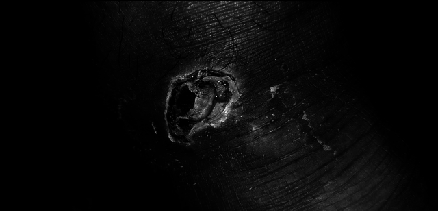} &
  \includegraphics[width=4cm,height=3.2cm]{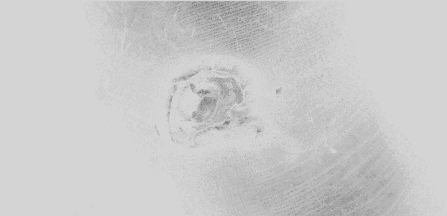} \\ \\

  \includegraphics[width=4cm,height=3.2cm]{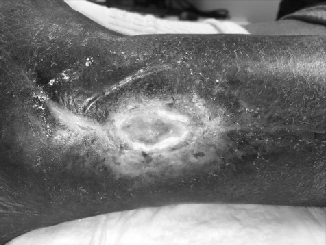} &
  \includegraphics[width=4cm,height=3.2cm]{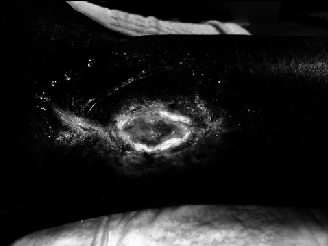} &
  \includegraphics[width=4cm,height=3.2cm]{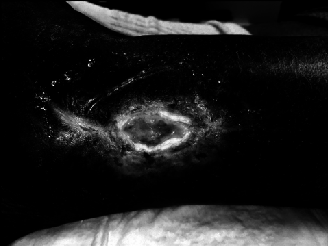} &
  \includegraphics[width=4cm,height=3.2cm]{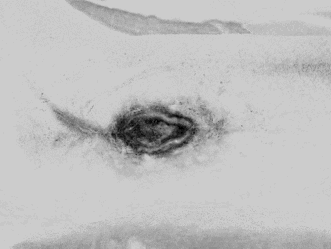} \\

  \end{tabular}
  \caption[]{Illustration of 2 cases from test set B showing the Y channel and its alternate representations. Y - luminance, eY - exaggerated luminance, eYS-R\&B - exaggerated luminance with swapped R and B coefficients. Note that the Difference images show the difference in features between the eY and eYS-R\&B images. The first row image is from the Alzubaidi dataset, and the second row image is from the FUSC dataset.}
  \label{fig:ey}
\end{figure*}

Table \ref{table:tensor_merging_ey_exponents} shows results for obtaining the optimum exponent value in the RGB+eY switched coefficient experiments. We initially selected an exponent value of 5, then experimented with values of 4 and 6. The results indicate that an exponent value of 5 provides the optimum exponent value. Figure \ref{fig:ey} shows two wound images from test set B for the luminance channel and the two alternate representations (eY and eY with swapped R and B coefficients). These images show a notable change in contrast between wound and non-wound regions. To the human eye, there is little discernible difference between eY and eYS-R\&B, although as shown in our results (see Table \ref{table:tensor_merging_ey_coef}) the latter offers test performance improvements over the former. The ``Difference" column shows the difference between the eY and eYS-R\&B channels, which indicates a dense concentration of features within the wound regions. The ``Difference" images were produced using the absdiff function in the Python CV2 library (\cite{opencv2000}). 

\begin{table*}[!h]
  \centering
  \caption{Summary of results for the multi-colour channel tensor merging operations when merging RGB colour channels with exaggerated luminance (eY) for switched R and B coefficients using different exponent (EX) values.}
  \label{table:tensor_merging_ey_exponents}
  \scalebox{0.77}{
  \begin{tabular}{|p{3.4cm}|p{1.7cm}|p{1.4cm}|p{1.5cm}|p{1.5cm}|p{1.2cm}|p{1.2cm}|p{1.3cm}|p{1.3cm}|p{1.4cm}|p{1.0cm}|p{1.0cm}|}
  \hline
  Colour Channels & Best Epoch & Train IoU & Train Loss & Train DSC & Val IoU & Val Loss & Val DSC & Test IoU & Test DSC & FPE    & FNE    \\ \hline \hline
  RGB (baseline)  & 50         & \textbf{0.9427}    & \textbf{0.0975}     & \textbf{0.9694}    & 0.6258  & 0.4063   & 0.7176  & 0.5350   & 0.6389   & 0.0597 & 0.3254 \\  
  RGB+eY (EX=4)   & 21         & 0.7383    & 0.3302     & 0.8405    & \textbf{0.6302}  & \textbf{0.3649}   & \textbf{0.7216}  & 0.5427   & 0.6468   & 0.0458 & 0.3190 \\
  RGB+eY (EX=5)   & 32         & 0.8585    & 0.2089     & 0.9213    & 0.6232  & 0.3903   & 0.7128  & \textbf{0.5576}   & \textbf{0.6654}   & \textbf{0.0420} & \textbf{0.3119} \\
  RGB+eY (EX=6)   & 31         & 0.8616    & 0.2041     & 0.9234    & 0.6288  & 0.3750   & 0.7184  & 0.5559   & 0.6630   & 0.0449 & 0.3387 \\
  \hline
  \end{tabular}
  }
\end{table*}

\subsubsection{Combined Instance-batch Normalisation}
The second of our modifications utilises a combined instance and batch normalisation (IBN) layer in the first convolution block of the encoder path. The IBN activation layer improves the ability of the encoder to extract features where contrast is still a prominent feature, found predominantly in the early layers of the encoder. 
In isolation, instance normalisation reduces contrast features but also reduces useful information, while batch normalisation allows for more of those features to be retained (\cite{pan2018ibn}). The procedure for creating an IBN layer is detailed in Algorithm \ref{alg:ibn}. 

\begin{algorithm}[!h]
  \caption{Instance-batch normalisation algorithm.}
  \label{alg:ibn}
  \begin{algorithmic}[1]
    \Procedure{IBN}{$channels$}
    \State $ratio \leftarrow 0.5$
	\State $half \leftarrow (channels \times ratio)$
    \State $in \leftarrow instance\_norm(half)$
    \State $bn \leftarrow batch\_norm(channels - half)$
    \State $out \leftarrow concatenate(in, bn)$
    
	\State Return $out$
	\EndProcedure
  \end{algorithmic}
\end{algorithm}


We experimented with IBN by gradually adding it to each successive convolution block in the encoder until performance started to degrade. Combining instance normalisation with batch normalisation ensures that the benefits of instance normalisation (removing contrast information (\cite{ulyanov2017instance})), are not lost while also benefiting from the effect of batch normalisation, which reduces internal covariate shift, stabilising training by reducing overfitting and improving model generalisation (\cite{ioffe2015batch}). The integration of batch normalisation ensures that the instance normalisation component does not remove more than the contrast features. 
This modification to HarDNet-DFUS is inspired by the work of \cite{pan2018ibn}. They demonstrated the effect of combining instance and batch normalisation in object classification and non-medical segmentation tasks. To the best of our knowledge, the use of IBN in our proposed HarDNet-CWS architecture is the first time that the method has been demonstrated in any deep learning chronic wound study.

\subsubsection{Parameterised Rectified Linear Unit}
The third adjustment we make to the HarDNet-DFUS architecture is the replacement of Rectified Linear Unit (ReLU) activation layers in the encoder convolution blocks with Parametric ReLU (PReLU) activation layers. PReLU is an advanced variation of prior ReLU activation functions (ReLU and Leaky ReLU) that has been shown to improve model fitting (\cite{he2015rectifiers}). PReLU can be used in training scenarios using backpropagation and can be optimised concurrently with other network layers. Leaky ReLU multiplies negative inputs by a nominal value, e.g. 0.022. PReLU improves on this aspect by making the nominal negative value learnable during training, allowing it to adapt more to weight and bias parameters. The mathematical expression for PReLU is show in Equation \ref{eq:prelu}. Conditionally, if $a_i = 0$, then $f$ becomes a ReLU activation. If $a_i > 0$, then $f$ becomes a leaky ReLU activation. If $a_i$ is learnable, then $f$ becomes a PReLU activation. 

\begin{equation}
f(y_i)= 
\begin{cases}
    {y_i},  & \text{if } y_i>0\\
    a_iy_i, & \text{if } y_i\leq 0
\end{cases}
\label{eq:prelu}
\end{equation}
\noindent
where $y_i$ is an input for the $i$th channel, and $a_i$ is the learnable parameter (negative slope). 

\subsubsection{Switchable-Normalisation}
To further enhance the encoder in our proposed network architecture, we implement a Switchable-Normalisation (SN) layer in the second encoder block. As with the previous experiments using IBN, we introduced SN into all layers of the encoder and gradually removed each layer, starting from the last layer, until the optimum performance was reached. SN, originally proposed by \cite{luo2021switchnorm}, selectively learns different normalisers by using channel, layer, and minibatch values to compute means and variance statistics. SN is able to adapt to various network architecture designs, is robust to a range of batch sizes, and is not prone to hyper-parameter sensitivity as exhibited by other normalisation methods such as group normalisation. SN inherits all the benefits of instance norm, layer norm, and batch norm by learning their importance ratios during training, preventing overfitting by balancing between generalisation and feature learning. The switchable-normalisation process is summarised in Equation \ref{eq:switch_norm}. 

\begin{equation}
  \Phi = \left \{ \lambda in, \lambda ln, \lambda bn, \lambda 'in, \lambda 'ln, \lambda 'bn \right \}
\label{eq:switch_norm}
\end{equation}
\noindent
where $\Phi$ is a set of learnable parameters, $in$ represents instance normalisation, $ln$ represents layer normalisation, and $bn$ represents batch normalisation. 

\begin{figure}[!h]
  \centering
  \begin{tabular}{cc}
    \includegraphics[width=3cm,height=5.5cm]{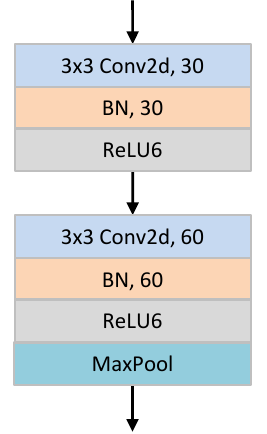} &
    \includegraphics[width=3cm,height=5.5cm]{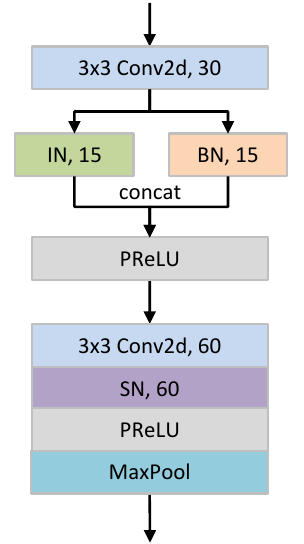} \\
    (a) & (b) \\
  \end{tabular}
  \caption[]{Illustration of (a) the original HarDNet-DFUS convolutional block design found in the encoder stem, and (b) our enhanced block design utilising instance-batch normalisation, PReLU activation, and Switchable Normalisation. BN - batch normalisaton, ReLU - rectified linear unit, IN - instance normalisation, PReLU - parametric rectified linear unit, SN - switchable normalisation.}
  \label{fig:hardnet_blocks}
\end{figure}

Figure \ref{fig:hardnet_blocks} shows the original block design for the stem layers of the HarDNet-DFUS encoder together with our proposed adjustments implementing IBN, PReLU, and SN. 

\subsubsection{Refined HarDNet Block Harmonic Structure}
\begin{figure}[!h]
  \centering
  \begin{tabular}{c} 
  \includegraphics[scale=.75]{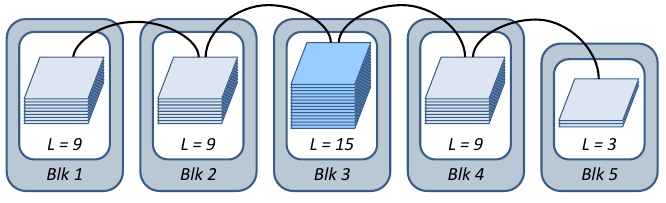} \\
  (a) \\ \\
  \includegraphics[scale=.75]{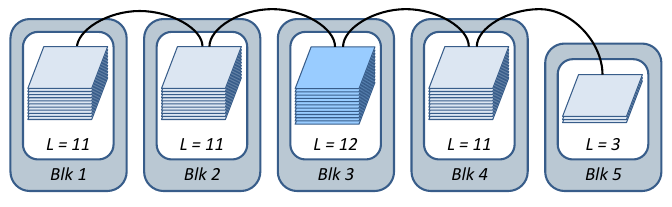} \\
  (b) \\ 
  \end{tabular}
  \caption{Illustration of (a) the original HarDNet-DFUS harmonic block design, and (b) our proposed HarDNet-CWS harmonic block which increases the density of the lower density blocks, and reduces the density of the higher density blocks which results in a reduction of the overall harmonic amplitude. L - number of layers in HarDNet convolution block.}
  \label{fig:harmonic}
\end{figure}

\begin{figure}[!h]
  \centering
  \begin{tabular}{c} 
  \includegraphics[scale=.48]{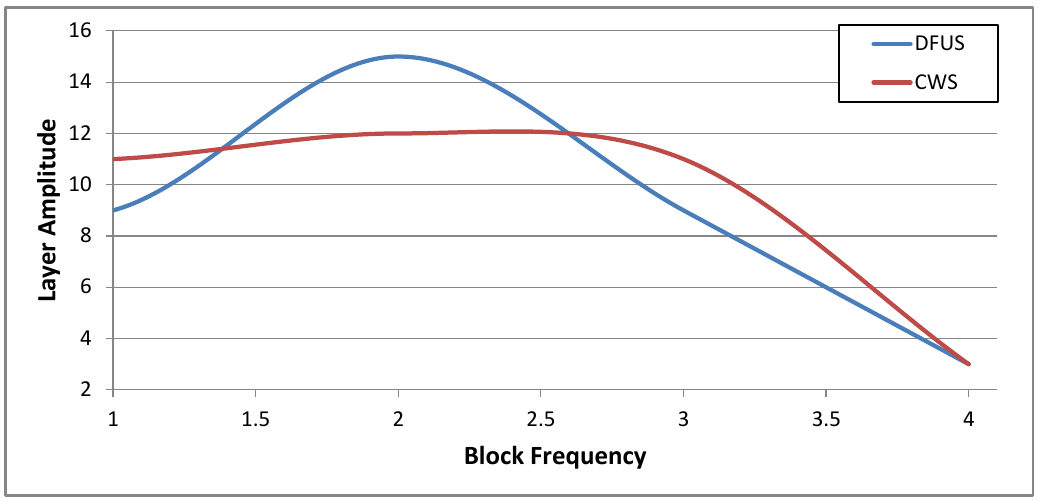}
  \end{tabular}
  \caption{Illustration comparing the waveform representations of the HarDNet blocks for the original HarDNet-DFUS architecture (blue), and our proposed HarDNet-CWS architecture (red).}
  \label{fig:hardnet_waveforms}
\end{figure}

The fourth refinement to our proposed HarDNet-CWS model architecture involves the adjustment of the harmonic shape found in the HarDNet convolution encoder blocks. The original block design is a pattern of increasing and decreasing sequence of convolution layers represented by each HarDNet block. In our proposed adjustment to the HarDNet blocks, we change the harmonic pattern such that the minimum and maximum layer amplitude values for the first four blocks are less pronounced. For the first four HarDNet blocks the number of layers in blocks with lower layer counts are increased, while the blocks with higher layer counts are reduced, creating a smoother harmonic pattern. This also results in an overall increase in distributed layers to facilitate the supplemental features captured from the additional eY channel tensors. Figure \ref{fig:harmonic} shows the original harmonic block design (a), and our improved harmonic block design (b). Figure \ref{fig:hardnet_waveforms} shows a comparison of the block and layer patterns expressed as waveforms for the original HarDNet-DFUS and our proposed HarDNet-CWS architecture. Our experimental results indicated that the network architecture responds more to lower variations in layer counts for each HarDNet block in the encoder when trained, validated, and tested on our wound datasets. The layer amplitude for HarDNet-DFUS has a $sd = 4.2427$, while our proposed HarDNet-CWS has a layer amplitude with $sd = 3.7149$. 


\begin{figure*}[!h]
  \centering
  \begin{tabular}{c}
    \includegraphics[scale=.71]{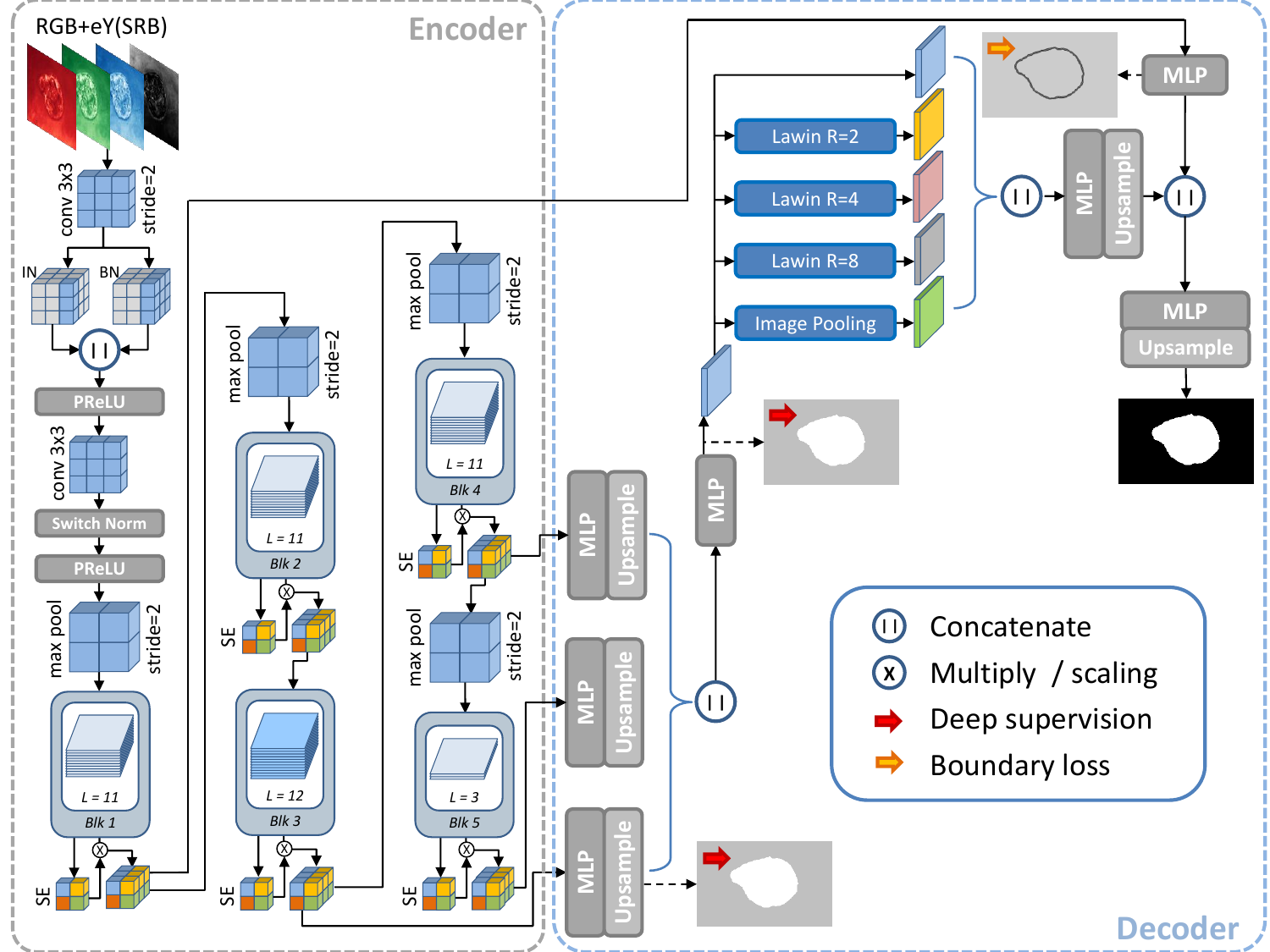} \\
  \end{tabular}
  \caption[]{Illustration of the proposed HarDNet-CWS network architecture. eY - exaggerated luminance, SRB - switched red and blue coefficients, IN - instance normalisation, BN - batch normalisation, SE - squeeze and excite, L - layers, Blk - HarDNet block, MLP - multilayer perceptron, R - patch size.}
  \label{fig:hardnet-cws}
\end{figure*}

Table \ref{table:improvements} shows a summary of all the proposed network architecture modifications. These results show that the highest performance increase is with the use of the CWS model trained using RGB+eY merged tensors with the proposed PReLu, IBN, SN, and HarDNet block harmonic adjustments. When using RGB+eY merged tensors with the proposed model adjustments, we observe test set performance improvements in terms of test IoU ($+0.0144$) and test DSC ($+0.0141$) when compared to using only RGB+eY merged tensors, as shown in the previous experiments. Figure \ref{fig:hardnet-cws} shows an overview of the proposed HarDNet-CWS architecture. 

\begin{table*}[!h]
  \centering
  \caption{Summary of results for the proposed model architecture improvements for HarDNet-CWS. DFUS - HarDNet-DFUS, CWS - HarDNet-CWS, eY - exaggerated luminance, IBN - instance-batch normalisation, PR - PReLU activation function, SN - switchable normalisation, Har - harmonic block adjustment.}
  \label{table:improvements}
  \scalebox{0.7}{
  \begin{tabular}{|p{6.42cm}|p{1.7cm}|p{1.4cm}|p{1.5cm}|p{1.5cm}|p{1.2cm}|p{1.2cm}|p{1.3cm}|p{1.3cm}|p{1.4cm}|p{1.0cm}|p{1.0cm}|}
  \hline
  Model                       & Best Epoch & Train IoU & Train Loss & Train DSC & Val IoU & Val Loss & Val DSC & Test IoU & Test DSC & FPE    & FNE    \\ \hline \hline
  DFUS (baseline)             & 50         & \textbf{0.9427}    & \textbf{0.0975}     & \textbf{0.9694}    & 0.6258  & 0.4063   & 0.7176  & 0.5350   & 0.6389   & 0.0597 & 0.3254 \\
  CWS RGB+Y+A         & 33         & 0.8745    & 0.1895     & 0.9310    & \textbf{0.6319}  & 0.3581   & \textbf{0.7229}  & 0.5530   & 0.6630   & 0.0508 & 0.3199 \\
  CWS [RGB+Y+A]+[PReLU+IBN+SN+Har] & 25    & 0.7903   & 0.2769  & 0.8769  & 0.6266 & \textbf{0.3572} & 0.7171 & 0.5570  & 0.6645 & \textbf{0.0417} & 0.3563 \\
  CWS RGB+eY                  & 32         & 0.8585    & 0.2089     & 0.9213    & 0.6232  & 0.3903   & 0.7128  & 0.5576   & 0.6654   & 0.0420 & \textbf{0.3119} \\
  CWS [RGB+eY]+[PReLU+IBN+SN+Har] & 27         & 0.8241    & 0.2483     & 0.9001    & 0.6193  & 0.3916   & 0.7082  & \textbf{0.5720}   & \textbf{0.6795}   & 0.0476 & 0.3132 \\
  \hline
  \end{tabular}
  }
\end{table*}

\subsection{GAN-based Pretraining}
\begin{figure}[!h]
  \centering
  \begin{tabular}{ccc}
  (a) & (b) & (c) \\
  \includegraphics[width=2.5cm,height=2.5cm]{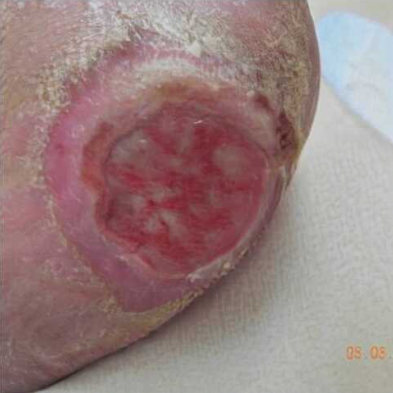} &
  \includegraphics[width=2.5cm,height=2.5cm]{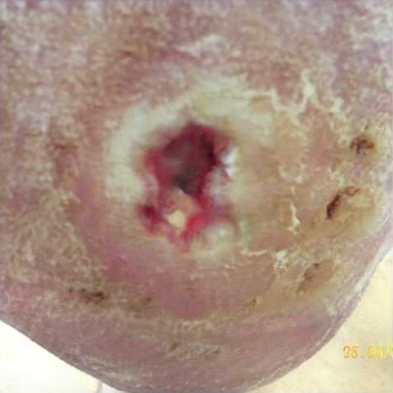} &
  \includegraphics[width=2.5cm,height=2.5cm]{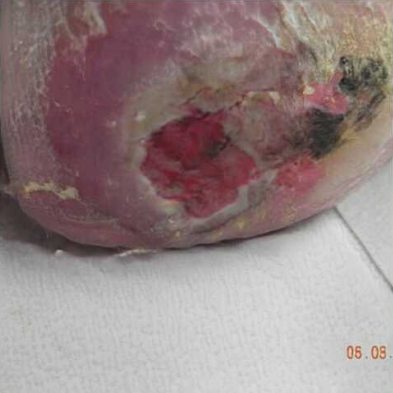} \\

  \end{tabular}
  \caption[]{Illustration of three GAN-generated DFU wounds from the 18,799 GAN-generated wound images that we used for pretraining our proposed HarDNet-CWS model.}
  \label{fig:gan_examples}
\end{figure}

\cite{alzubaidi2020medical} conducted experiments in DFU wound classification with different transfer learning scenarios. They showed that same-domain transfer learning significantly improved model performance. 
\cite{brungel2023unconditionally} would later conduct DFU segmentation experiments using 4000 GAN-generated DFU wound images to improve performance of a segmentation model. 
In this section we experiment with a model trained and validated on a solely synthetic DFU segmentation dataset, which we then use as a pretrained model for training our proposed HarDNet-CWS model. The respective dataset, consisting of 20,000 unconditionally generated and pseudo-labelled DFU images, originating from groundworks of \cite{brungel2023unconditionally} and provided for this study. Two underlying GAN-models were trained on the DFUC2022 dataset, one on the training set and one on the training and test set. From each, 10,000 images were generated via incrementing seeds and pseudo-labelled as described in the original work. Of these a total of 18,799 samples with at least one DFU instance was selected, and samples not showing any instances were discarded. Figure \ref{fig:gan_examples} shows a selection of images from the included samples, demonstrating the variety of generated representations. For model training we then split the synthetic dataset using an 80:20 ratio into a training set ($n = 15,039$) and validation set ($n = 3760$). We then trained our best model using this data. Next, we froze the stem layers and the first HarDNet block in our model, and trained again using the trained GAN DFU model as pretrained weights. The results of this experiment are shown in Table \ref{table:gan_train}. 
When compared to the best performing model from the previous experiments (CWS+[RGB+eY]+[PReLU+IBN+SN+Har]), the results for the test set show clear performance improvements in terms of test IoU ($+0.0243$), test DSC ($+0.0212$), FPE ($-0.0032$), and FNE ($-0.0032$). 
 
\begin{table*}[!h]
  \centering
  \caption{Summary of results showing the performance improvements for the proposed HarDNet-CWS model when using the DFU GAN pretrained weights. CWS - HarDNet-CWS [RGB+eY]+[PReLU+IBN+SN+Har].}
  \label{table:gan_train}
  \scalebox{0.82}{
  \begin{tabular}{|p{2.5cm}|p{1.65cm}|p{1.4cm}|p{1.5cm}|p{1.5cm}|p{1.2cm}|p{1.2cm}|p{1.3cm}|p{1.3cm}|p{1.4cm}|p{1.0cm}|p{1.0cm}|}
  \hline
  Model                 & Best Epoch & Train IoU & Train Loss & Train DSC & Val IoU & Val Loss & Val DSC & Test IoU & Test DSC & FPE    & FNE    \\ \hline \hline
  DFUS (baseline)             & 50         & 0.9427    & 0.0975     & 0.9694    & 0.6258  & 0.4063   & 0.7176  & 0.5350   & 0.6389   & 0.0597 & 0.3254 \\
  CWS & 27         & 0.8241    & 0.2483     & 0.9001    & 0.6193  & 0.3916   & 0.7082  & 0.5720   & 0.6795   & 0.0476 & 0.3132 \\
  CWS+pretrained        & 40  & \textbf{0.9444}    & \textbf{0.0961}     & \textbf{0.9704}    & \textbf{0.6713}  & \textbf{0.3391}   & \textbf{0.7580}  & \textbf{0.5963}   & \textbf{0.7007}   & \textbf{0.0444} & \textbf{0.3100} \\
  \hline
  \end{tabular}
  }
\end{table*}

\subsection{Cross-domain Weakly Supervised Training Using Animal Meat Dataset}
\begin{figure}[!h]
  \centering
  \begin{tabular}{ccc}
  (a) & (b) & (c) \\
  \includegraphics[width=2.5cm,height=2.5cm]{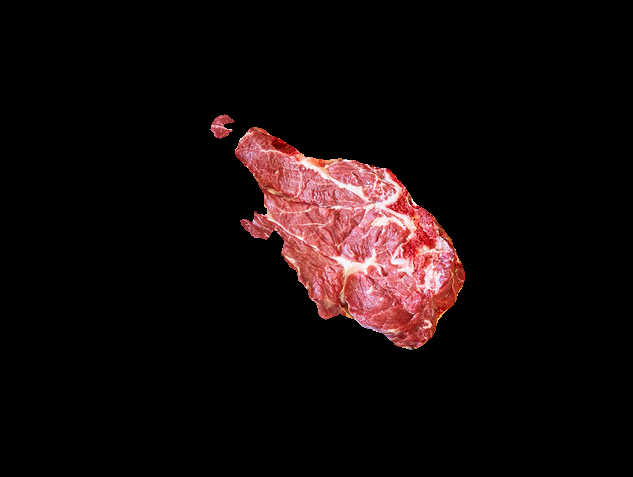} &
  \includegraphics[width=2.5cm,height=2.5cm]{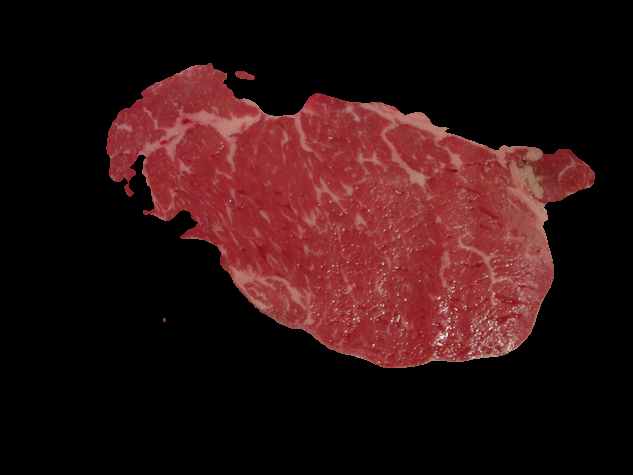} &
  \includegraphics[width=2.5cm,height=2.5cm]{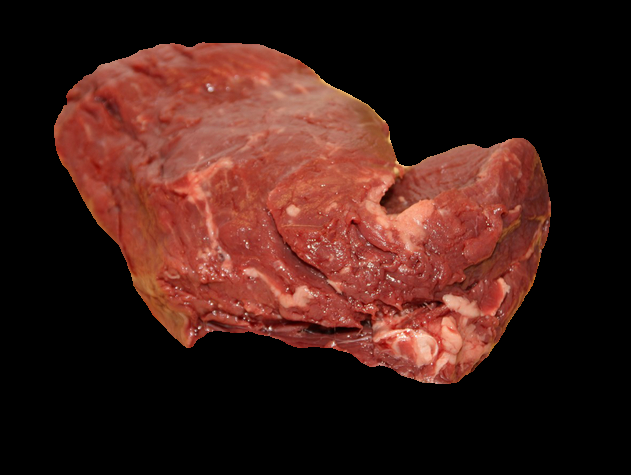} \\

  \end{tabular}
  \caption[]{Illustration of three masked animal meat images used in the weakly supervised training process to enhance performance of our HarDNet-CWS model. Prior to training, ground truth masks were generated via inference using our best model.}
  \label{fig:meat_examples}
\end{figure}

In this experiment, we sourced a dataset of 363 animal meat images using Google Image Search with the Creative Commons License search option to remove copyrighted images from search results. The motivation for this experiment derives from the visual appearance of textures present in both cooked and uncooked animal meat, which we identified as being similar to those of human wounds. Given the small size of the animal meat dataset, rather than using pretraining, we include the images directly into the wound training set. Beforehand, we used our current best model to complete inference on the animal meat images and used the resulting prediction masks as ground truth. Table \ref{table:meat_results} shows the results of the experiments which introduced the animal meat dataset into the training workflow. When compared to the best performing model in the previous experiments (CWS+pretrained), these results show clear performance improvements for test IoU ($+0.0138$), test DSC ($+0.0154$), and FNE ($-0.0109$). Figure \ref{fig:meat_examples} shows three example masked animal meat images that we used to enhance model performance. 

\begin{table*}[!h]
  \centering
  \caption{Summary of results showing the performance improvements when introducing the animal meat dataset into the training process. BEp - best epoch, CWS - HarDNet-CWS [RGB+eY]+[PReLU+IBN+SN+Har], AMD - animal meat dataset.}
  \label{table:meat_results}
  \scalebox{0.82}{
  \begin{tabular}{|p{3.6cm}|p{0.7cm}|p{1.4cm}|p{1.5cm}|p{1.5cm}|p{1.2cm}|p{1.2cm}|p{1.3cm}|p{1.3cm}|p{1.4cm}|p{1.0cm}|p{1.0cm}|}
  \hline
  Model                 & BEp & Train IoU & Train Loss & Train DSC & Val IoU & Val Loss & Val DSC & Test IoU & Test DSC & FPE    & FNE    \\ \hline \hline
  DFUS (baseline)             & 50         & 0.9427    & 0.0975     & 0.9694    & 0.6258  & 0.4063   & 0.7176  & 0.5350   & 0.6389   & 0.0597 & 0.3254 \\
  CWS+pretrained        & 40  & 0.9444    & 0.0961     & 0.9704    & 0.6713  & 0.3391   & 0.7580  & 0.5963   & 0.7007   & \textbf{0.0444} & 0.3100 \\
  CWS+pretrained+AMD & 52  & \textbf{0.9509} & \textbf{0.0857}     & \textbf{0.9738}    & \textbf{0.6759}  & \textbf{0.3213}   & \textbf{0.7660}  & \textbf{0.6101}   & \textbf{0.7161}   & 0.0456 & \textbf{0.2991}  \\
  \hline
  \end{tabular}
  }
\end{table*}

\subsection{Augmentation and K-Fold Cross Validation}
For the final stage of training our proposed HarDNet-CWS model, we completed a 5-fold cross validation together with training augmentation and test time augmentation (TTA) to enhance model performance. 
For the training augmentation, the albumentations library (\cite{buslaev2020albumentations}) was utilised to generate the following: 
(1) center cropping; (2) random cropping; (3) horizontal flipping; (4) vertical flipping; (5) shift scale with rotation; (6) Gaussian noise; (7) random brightness and contrast; (8) contrast limited adaptive histogram equalisation; and (9) multi-scaling. For TTA we employed horizontal and vertical flipping. 
The training and validation results for these experiments are summarised in Table \ref{table:5fold_train_val}. When compared to the best performing model from the previous experiments (CWS+PT+AMD), these results show clear performance improvements on the test set for the CWS+PT+AMD+5F+TTA model in terms of test IoU ($+0.0519$), test DSC ($+0.0449$), and FNE ($-0.0489$). 

\begin{table*}[!h]
  \centering
  \caption{Summary of results showing the performance improvements when using 5-fold cross validation (5F) and test time augmentation (TTA). BEp - best epoch, CWS - HarDNet-CWS [RGB+eY]+[PReLU+IBN+SN+Har], PT - pretrained, AMD - animal meat dataset.}
  \label{table:5fold_train_val}
  \scalebox{0.81}{
  \begin{tabular}{|p{3.83cm}|p{0.7cm}|p{1.4cm}|p{1.5cm}|p{1.5cm}|p{1.2cm}|p{1.2cm}|p{1.3cm}|p{1.3cm}|p{1.4cm}|p{1.0cm}|p{1.0cm}|}
  \hline
  Model                 & BEp & Train IoU & Train Loss & Train DSC & Val IoU & Val Loss & Val DSC & Test IoU & Test DSC & FPE    & FNE    \\ \hline \hline
  DFUS (baseline)             & 50         & 0.9427    & 0.0975     & 0.9694    & 0.6258  & 0.4063   & 0.7176  & 0.5350   & 0.6389   & 0.0597 & 0.3254 \\
  
  CWS+PT+AMD & 52  & \textbf{0.9509} & \textbf{0.0857}     & \textbf{0.9738}    & 0.6759  & \textbf{0.3213}   & 0.7660  & 0.6101   & 0.7161   & \textbf{0.0456} & 0.2991  \\
  
  CWS+PT+AMD+5F     & 59 & 0.7561 & 0.3049 & 0.8507 & 0.6822 & 0.3571 & 0.7775 & 0.6460 & 0.7485 & 0.0526 & 0.2672 \\
  
  CWS+PT+AMD+5F+TTA & 59 & 0.7561 & 0.3049 & 0.8507 & \textbf{0.6822} & 0.3571 & \textbf{0.7775} & \textbf{0.6620} & \textbf{0.7610} & 0.0522 & \textbf{0.2502} \\
  \hline
  \end{tabular}
  }
\end{table*}

\subsection{Qualitative Analysis}
Two clinical experts from two different hospitals were recruited, each with more than 10 years clinical experience, to rate the inference predictions from the HarDNet-DFUS (baseline) and HarDNet-CWS (proposed) models for test sets A and B using a 5-star rating system. A rating of 1 indicates a poor quality prediction, while a rating of 5 indicates an excellent quality prediction. Raters were asked to not rate a prediction if the model failed to make any prediction where wounds were visible in the image. If no wounds were present in an image and no prediction had been generated, then raters were asked to rate the prediction with a 5-rating. If more than one wound was present in an image, then the raters were asked to rate the overall quality of all predictions in the image. To reduce possible bias, raters were not informed of which model prediction images came from. 

Statistical analysis to ascertain reliability measures taken from two clinical experts who rated the HarDNet-DFUS (baseline) and HarDNet-CWS (proposed) test results was completed using IBM SPSS version 28.0.1.0 (SPSS Inc., Chicago, Illinois). The analysis of the ordinal data was completed using the intra-class correlation coefficient (ICC) to obtain inter-rater reliability consistency and agreement measures. 
Consistency is defined as the degree to which the score of a single rater ($y$) can be equated to a second rater's score ($x$) plus a systematic error ($c$) (i.e., $y = x + c$). 
Agreement concerns the extent to which $y$ is equal to $x$ (\cite{koo2016icc}). 
A two-way random effects model was used to generalise results to a population of raters from which the clinical expert raters in our study represent a sample. 
The mathematical expressions for ICC consistency and ICC agreement are shown in Equations \ref{eq:icc_con} and \ref{eq:icc_agree} respectively.
\begin{align}
  \begin{split}
    ICC &= \frac{MS_R-MS_E}{MS_R}
    \label{eq:icc_con}
  \end{split}
  \intertext{}
  \begin{split}
    ICC &= \frac{MS_R-MS_E}{MS_R+\frac{MS_C-MS_E}{n}}
    \label{eq:icc_agree}
  \end{split}
\end{align}

\noindent where $MS_R$ is the mean square for rows, $MS_E$ is the mean square for error, $MS_C$ is the mean square for columns, and $n$ is the number of subjects. 

ICC values are interpreted as follows: 0-0.39 indicates poor reliability; 0.4-0.74 indicates moderate reliability; 0.75-1 indicates excellent reliability (\cite{fleiss1999clinical}).

\section{Results}
In this section we report on the results of inference using our proposed HarDNet-CWS model. We present the results for two test sets: test set A which comprises 342 dark skin tone wound images and corresponding masks taken from the DFUC2022, AZH, CWDB, and FUSC datasets; and test set B which comprises 342 dark skin tone wound images with no masks taken from the Alzubaidi, Fitzpatrick17k, FUSC, GIS-W, Medetec, Wseg, and KSUMC datasets. The test set A predictions were assessed quantitatively and qualitatively, and the test set B results were assessed qualitatively only as this test set has no ground truth masks. 

\subsection{Quantitative Results for Test Set A}
\begin{figure}[!h]
  \centering
  \begin{tabular}{ccc}
  (Wound Image) & (DFUS) & (CWS) \\
  \includegraphics[width=2.4cm,height=3.4cm]{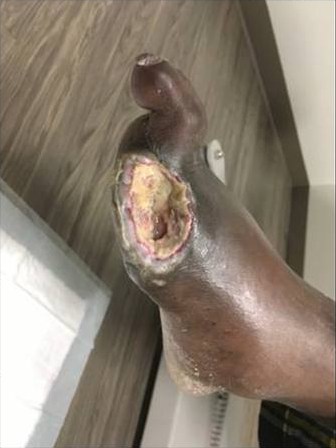} &
  \includegraphics[width=2.4cm,height=3.4cm]{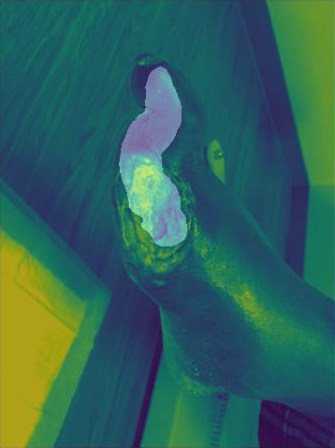} &
  \includegraphics[width=2.4cm,height=3.4cm]{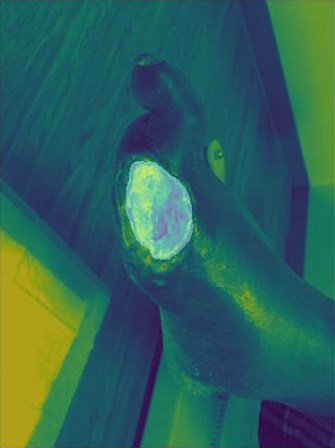} \\

  \includegraphics[width=2.4cm,height=3.4cm]{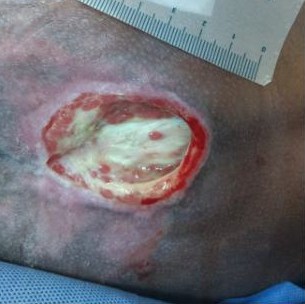} &
  \includegraphics[width=2.4cm,height=3.4cm]{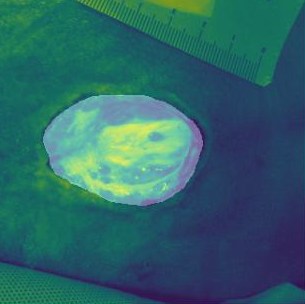} &
  \includegraphics[width=2.4cm,height=3.4cm]{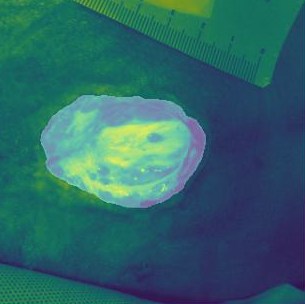} \\

  \includegraphics[width=2.4cm,height=3.4cm]{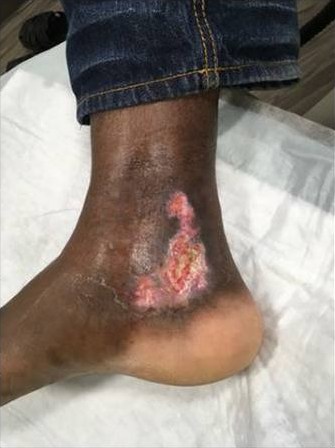} &
  \includegraphics[width=2.4cm,height=3.4cm]{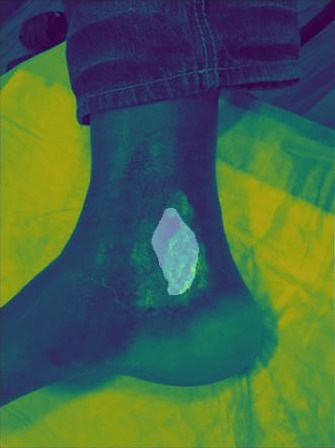} &
  \includegraphics[width=2.4cm,height=3.4cm]{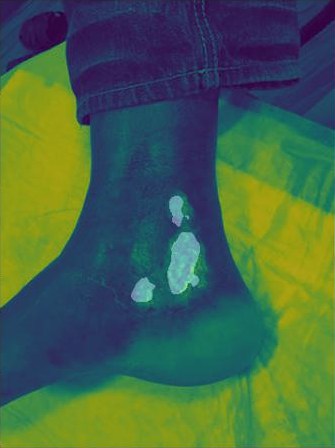} \\
  \end{tabular}
  \caption[]{Illustration of a selection of wound segmentation predictions from test set A for the HarDNet-DFUS (baseline) and HarDNet-CWS (proposed) models. The first row shows a DFU wound on a foot exhibiting partial amputation, the second row shows a PRU wound on the lower back of the torso, and the third row shows a DFU wound on the ankle. The first and third row images are from the FUSC dataset, and the second row image is from the CWDB dataset.}
  \label{fig:test_set_a_example_results}
\end{figure}

Test metrics for test set A inference results for the HarDNet-DFUS (baseline) and HarDNet-CWS (proposed) models are summarised in Table \ref{table:test_set_a_results}. We observe significant improvements in terms of IoU ($+0.1274$), DSC ($+0.1221$), and FNE ($-0.0752$), while FPE demonstrated a more subtle improvement ($-0.0075$). Figure \ref{fig:test_set_a_example_results} shows a selection of predictions from test set A demonstrating clear improvements in segmentation performance when comparing the baseline results from the HarDNet-DFUS model with the proposed HarDNet-CWS model. The first row shows a DFU wound on a foot exhibiting partial amputation, and shows that skin which has been miss-detected along the side of the toe with the DFUS model has not been inaccurately detected by the CWS model. This DFUS miss-detection may have been due to the darker skin on the toe, compared to the skin on the rest of the foot, which the model may have partly miss-detected as necrotic tissue. The second row shows a PRU wound on the lower-back of the torso where the CWS model has more accurately detected the edge details of the wound when compared to the DFUS prediction. This may be a result of the additional features provided by the enhanced tensor inputs in the CWS model, allowing the edge loss function to more accurately define wound boundary details. The third row shows a DFU wound on the ankle where the DFUS model prediction is more generalised and includes a significant region of miss-detected skin, and is much less accurate when compared to the CWS prediction. 

\begin{table}[!h]
    \centering
    \caption{Test results for the HarDNet-DFUS (baseline) and HarDNet-CWS (proposed) models for test set A dark skin tone wound images that have ground truth masks.}
    \label{table:test_set_a_results}
    \scalebox{1.0}{
    \begin{tabular}{|p{2.4cm}|p{1.0cm}|p{1.0cm}|p{1.0cm}|p{1.0cm}|}
    \hline
	  Model        & IoU    & DSC    & FPE    & FNE    \\ \hline \hline
    HarDNet-DFUS & 0.5350 & 0.6389 & 0.0597 & 0.3254 \\ 
    HarDNet-CWS  & \textbf{0.6624} & \textbf{0.7610} & \textbf{0.0522} & \textbf{0.2502} \\ 
    \hline
	\end{tabular}
	}
\end{table}

\subsection{Qualitative Results for Test Sets A and B}
Qualitative measures for test set A and B inference results from the HarDNet-DFUS (baseline) model and HarDNet-CWS (proposed) model are shown in Table \ref{table:test_a_ratings}. 
The ICC confidence and agreement values for the HarDNet-DFUS test set A predictions (confidence ICC = 0.6714, agreement ICC = 0.6717) indicate moderate reliability for the clinical ratings for this model. The ICC confidence and agreement values for the HarDNet-DFUS predictions for test set B (confidence ICC = 0.7907, agreement ICC = 0.7747) indicate excellent reliability for the clinical ratings for this model. 
The ICC confidence and agreement values for the HarDNet-CWS test set A predictions (confidence ICC = 0.6633, agreement ICC = 0.6631) indicate moderate reliability. 
The ICC confidence and agreement values for the HarDNet-CWS predictions for test set B (confidence ICC = 0.5001, agreement ICC = 0.4992) indicate moderate reliability. 
Overall, the ICC reliability measures for the DFUS (baseline) model predictions indicate moderate to excellent reliability, while moderate reliability is demonstrated for the CWS (proposed) model. 
For the CWS ICC test set A reliability measures, 311 ratings exactly matched, while 19 ratings varied by 1. 
For the CWS ICC test set B reliability measures, 308 ratings exactly matched, while 20 ratings varied by 1. These results indicate that the majority of ratings between raters matched exactly, or had a difference of no more than 1. 

\begin{table}[!h]
  \centering
  \renewcommand{\arraystretch}{1.0}
  \caption{Measures derived from expert rater quality assessment of test sets A and B inference results for the HarDNet-DFUS (baseline) and HarDNet-CWS (proposed) model. ICC - intra-class correlation coefficient, Co - consistency, Ag - agreement, LB - lower bound, UB - upper bound, CI - confidence interval.}
  \scalebox{0.89}{
    \label{table:test_a_ratings}
    \begin{tabular}{|c|c|c|c|c|c|}
    \hline
    Test Set & Seg Model & Type & ICC    & LB95\%CI & UB95\%CI \\ \hline \hline
    A        & DFUS      & Co   & 0.6714 & 0.5935   & 0.7343   \\ 
    A        & DFUS      & Ag   & 0.6717 & 0.5940   & 0.7346   \\ 
    B        & DFUS      & Co   & 0.7907 & 0.7411   & 0.8308   \\ 
    B        & DFUS      & Ag   & 0.7749 & 0.6986   & 0.8287   \\ 

    A        & CWS       & Co   & 0.6633 & 0.5835   & 0.7278   \\ 
    A        & CWS       & Ag   & 0.6631 & 0.5834   & 0.7276   \\ 
    B        & CWS       & Co   & 0.5001 & 0.3817   & 0.5959   \\ 
    B        & CWS       & Ag   & 0.4992 & 0.3809   & 0.5949   \\ 
    \hline
    \end{tabular}
  }
\end{table}

\begin{figure}[t!]
  \centering
  \begin{tabular}{c}
  Test Set A (Rater 1 and 2) \\
  \includegraphics[scale=0.33]{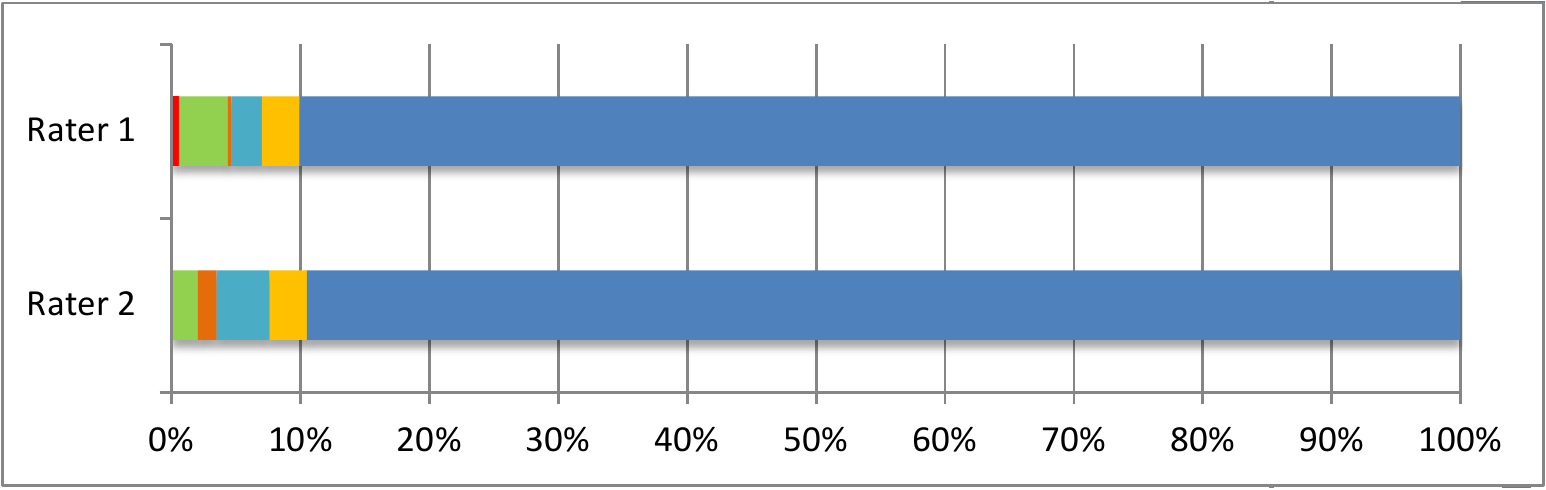} \\ \\

  Test Set B (Rater 1 and 2) \\
  \includegraphics[scale=0.495]{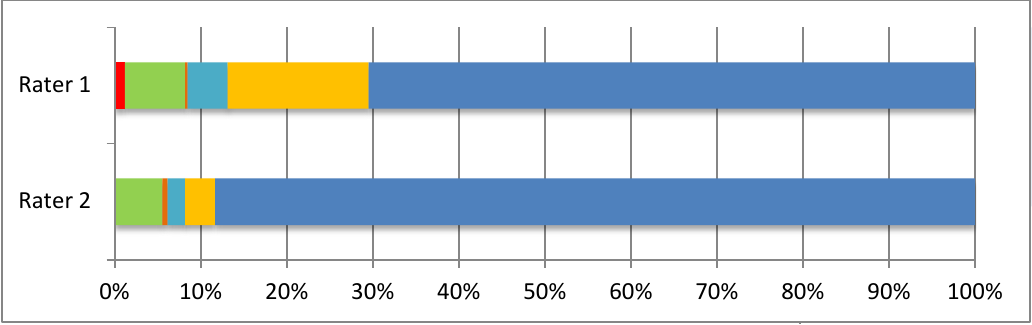} \\ \\
  
  Test Sets A and B (Rater 1 and 2) \\
  \includegraphics[scale=0.495]{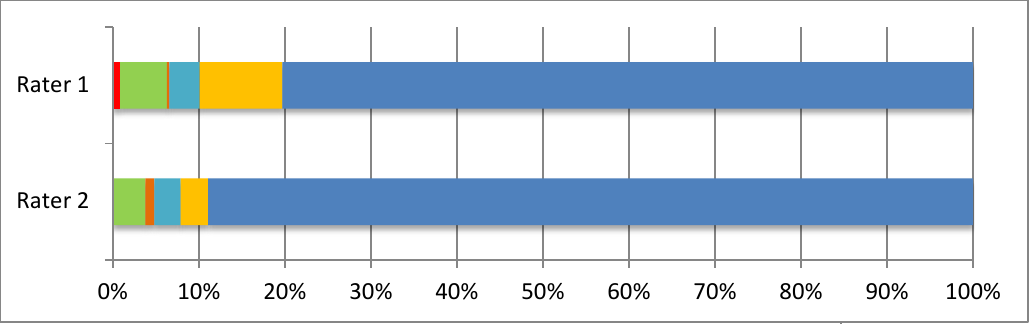} \\
  
  \includegraphics[scale=0.66]{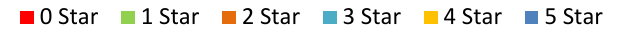}
  
  \end{tabular}
  \caption[]{Relative distribution of clinical ratings for test sets A and B DFUS (baseline) model predictions.}
  \label{fig:dfus_dist}
\end{figure}

To provide further insights into the clinician prediction ratings, we conducted a relative distribution analysis. A summary of the distribution analysis for the DFUS predictions is shown in Figure \ref{fig:dfus_dist}. These results indicate that for the DFUS (baseline) results, both raters consistently rated the predictions highly, within the 4-5 star range: 
test set A for rater 1 = 92.98\%, 
test set A for rater 2 = 92.39\%, 
test set B for rater 1 = 86.84\%, 
test set B for rater 2 = 91.81\%, 
test sets A and B for rater 1 = 89.91\%, and test sets A and B for rater 2 = 92.11\%. 

\begin{figure}[t!]
  \centering
  \begin{tabular}{c}
  Test Set A (Rater 1 and 2) \\
  \includegraphics[scale=0.495]{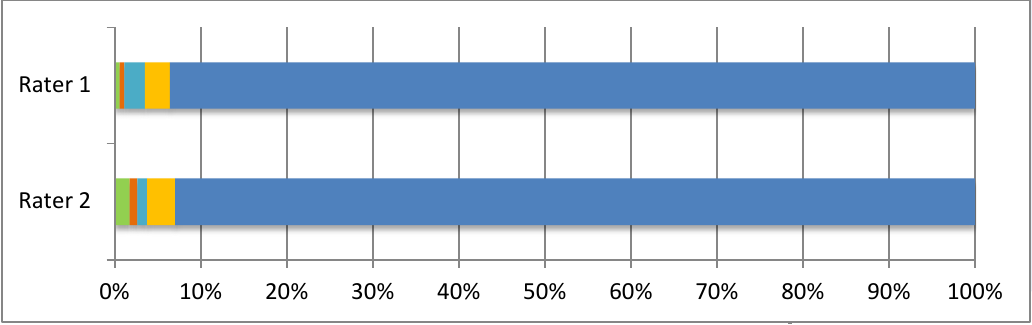} \\ \\

  Test Set B (Rater 1 and 2) \\
  \includegraphics[scale=0.495]{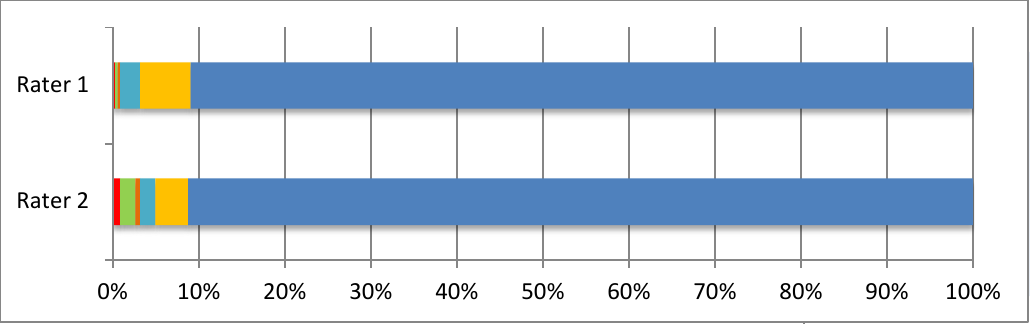} \\ \\
  
  Test Sets A and B (Rater 1 and 2) \\
  \includegraphics[scale=0.495]{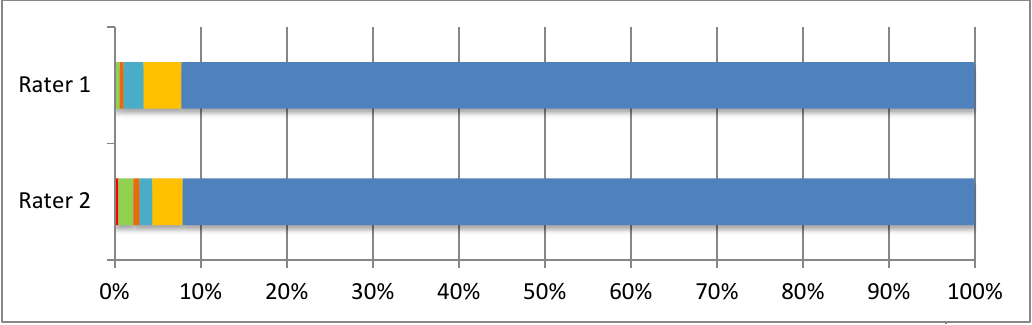} \\
  
  \includegraphics[scale=0.66]{rater_legend.pdf}
  
  \end{tabular}
  \caption[]{Relative distribution of clinical ratings for test sets A and B CWS (proposed) model predictions.}
  \label{fig:cws_dist}
\end{figure}

The distribution analysis for the CWS predictions is shown in Figure \ref{fig:cws_dist}. 
These results indicate that for the CWS (proposed) model, both raters consistently rated the predictions highly, within the 4-5 star range: 
test set A for rater 1 = 96.49\%, test set A for rater 2 = 96.20\%, 
test set B for rater 1 = 96.79\%, test set B for rater 2 = 95.87\%, 
test sets A and B for rater 1 = 96.64\%, 
and test sets A and B for rater 2 = 95.62\%. 

We observe that for both test sets and both raters, the CWS (proposed) predictions demonstrated higher scores than the DFUS (baseline) predictions in terms of expert qualitative assessment. A summary of the improvements demonstrated by the CWS (proposed) model based on expert qualitative assessment is shown in Table \ref{table:qual_improve} for 5 star ratings, and Table \ref{table:qual_improve_extra} for 4-5 star ratings. We observe that the number of 5 star ratings for rater 1 on test set B is significantly lower than the other 5 star ratings for this model. However, as shown in Table \ref{table:qual_improve_extra}, the difference is much less pronounced when taking into account 4-5 star ratings, meaning that the discrepancy is mostly due to a difference of 1 star between raters. 

\begin{table}[!h]
  \centering
  \renewcommand{\arraystretch}{1.0}
  \caption{Summary of percentage improvements in terms of 5 star ratings for the HarDNet-CWS (proposed) model when compared to the HarDNet-DFUS (baseline) model.}
  \scalebox{0.89}{
    \label{table:qual_improve}
    \begin{tabular}{|c|c|c|c|c|}
    \hline
    Test Set & Rater & DFUS 5 Star & CWS 5 Star & Improvement \% \\ \hline \hline
    A        & 1     & 90.06\%     & 93.57\%    & 3.51\%         \\
    A        & 2     & 89.47\%     & 92.98\%    & 3.51\%         \\
    B        & 1     & 70.47\%     & 90.94\%    & 20.47\%        \\    
    B        & 2     & 88.30\%     & 92.04\%    & 3.74\%        \\
    \hline
    \end{tabular}
  }
\end{table}

\begin{table}[!h]
  \centering
  \renewcommand{\arraystretch}{1.0}
  \caption{Summary of percentage improvements in terms of 4-5 star ratings for the HarDNet-CWS (proposed) model when compared to the HarDNet-DFUS (baseline) model.}
  \scalebox{0.84}{
    \label{table:qual_improve_extra}
    \begin{tabular}{|c|c|c|c|c|}
    \hline
    Test Set & Rater & DFUS 4-5 Star & CWS 4-5 Star & Improvement \% \\ \hline \hline
    A        & 1     & 92.98\%       & 96.49\%      & 3.51\%         \\
    A        & 2     & 92.39\%       & 96.20\%      & 3.81\%         \\
    B        & 1     & 86.84\%       & 96.79\%      & 9.95\%         \\
    B        & 2     & 91.81\%       & 95.87\%      & 4.06\%         \\
    \hline
    \end{tabular}
  }
\end{table}

\subsection{Test Set Images with Blank Masks}
During testing with test set A, we observed a number of cases where the ground truth masks comprised only of black pixels, indicating that there were no wound regions present in the corresponding images. However, qualitative results obtained from clinicians showed that some of these cases had in fact been labelled incorrectly. We identified 14 cases in test set A that were sourced from the AZH ($n = 4$), FUSC ($n = 9$), and DFUC2022 ($n = 1$) datasets where wounds were clearly present in the images, but the corresponding masks comprised of only black pixels. The total number of incorrectly labelled blank masks represents $\approx 4\%$ of a test set total (342 images / masks), indicating that the reported metrics in Tables 9 to 15 are likely to be under-estimates. 




\section{Discussion}
This work focuses primarily on subjective measures derived from expert assessment of model predictions - a facet which is absent from almost all chronic wound deep learning research. 
Our experiment results indicate significant disparities between the quantitative lab based results and the qualitative results obtained from clinical expert ratings for both baseline (HarDNet-DFUS) and proposed (HarDNet-CWS) models. 
However, the results for our proposed HarDNet-CWS model show clear performance improvements in terms of lab based metrics and expert qualitative assessment. 

The reliability measures obtained from both clinical expert raters for test sets A and B indicate that reliability is moderate to excellent for the baseline model, and is moderate for the proposed model. However, a further analysis of these results shows that for the proposed model, 311 of 342 5 star ratings matched between raters for test set A, and 308 of 342 5 star ratings matched between raters for test set B. Further, 19 ratings for test set A varied by only 1 star, and for test set B 20 ratings varied by only 1 star. For test set A, a total of 330 of 342 ratings either matched or differed by only 1 star, and for test set B a total of 328 of 342 ratings matched or differed by only 1 star. We therefore suggest that when taking into account that the majority of expert ratings ($> 95\%$) either matched or differed by only 1 star, these results should be considered to demonstrate generally excellent levels of agreement. 

Our proposed model was trained and validated on chronic wound images taken from patients with lighter skin, while the two test sets comprised only wound images acquired from patients with darker skin tones. We observe that the validation results for our best performing model on test set A (CWS+PT+AMD+5F+TTA - see Table \ref{table:5fold_train_val}) are marginally higher when compared to the IoU and DSC test results: +0.0202 val IoU compared to test IoU, +0.0165 val DSC compared to test DSC. These results may be evidence that models trained only on lighter skin wound images may find inference challenging on darker skin wound images. However, in the absence of qualitative comparisons between the validation and test inference results, and taking into account the significant disparity between the lab based metrics and the expert qualitative results, we suggest that the differences in validation (lighter skin) and test (darker skin) results may not provide a complete assessment of the model's true ability. 

A limitation of this work is that the lab based metrics are assessed on a more fine-grained continuous scale (0-0.1), while the qualitative measures are measured on a 0-5 star ordinal scale. Future work might focus on a more fine-grained approach to qualitative measures, although we suggest that our results give a good general indication of the qualitative aspects of model predictions. 

The colour aspects of deep learning research involving the use of medical colour imaging is relatively under-explored. Colour imaging provides an enhanced visualisation of dermatological surface and subsurface structures which present novel challenges. This is especially pertinent in the deep learning domain, as most methods focus on single-channel images, which are generally less applicable to multi-colour channel domains (\cite{celebi2022editorial}). In this paper, we make an attempt to direct focus on this aspect with the use of manipulated multi-colour space tensors and a corresponding modified hybrid transformer network architecture that facilitates the additional colour information. Our experiments seem to indicate that there may be additional features in different colour spaces, which the model is able to learn from when such colour space data is merged into single tensors. Our future work will continue to explore the colour aspects of medical wound photographs when training deep learning models. 

Our results indicate that there may be a limited capacity for lab-based accuracy metrics when using the current publicly available datasets. We posit that this is largely due to variability in segmentation labelling. This is especially pertinent in the case of chronic wound labelling, which has been shown to be highly variable and subjective (\cite{ramachandram2021seg}). The observed disparity between DSC / IoU and expert subjective ratings for model predictions in our study indicates that the lab-based metrics are only providing part of the picture in deep learning assessment. 

Recent studies, such as those conducted by \cite{combalia2022isic}, have highlighted a disparity in laboratory results obtained from deep learning models and results obtained in real-world scenarios. To address this issue, our study has an increased emphasis on presenting results from a qualitative analysis of the model predictions obtained in our wound segmentation experiments. The measures derived from our qualitative analysis clearly show that clinician ratings of model predictions are significantly more favourable when compared to the lab-based metrics. 

The test sets we used in our main experiments, comprising only darker skin tones, were relatively small compared to most test sets used in deep learning studies. However, this limitation is due to the number of publicly available chronic wound images with ground truth masks, and the limited available time of our clinical collaborators who provided the expert assessment of model predictions. Despite these limitations, the present work presents the most extensive qualitative study so far in chronic wound segmentation. 



This work represents the first study to identify that animal meat images can be used to enhance the performance of a chronic wound segmentation model. Using just 363 animal meat images, with weak supervision, we were able to improve model performance by 0.0141 for test DSC and 0.0144 for test IoU. Animal meat images are significantly easier to obtain than chronic wound images, and require no ethical approval to collect. Furthermore, it may be of interest to experiment with GANs that can generate additional meat images, and to experiment to see how much further such images can be used to boost chronic wound model performance. The number of publicly available chronic wound images with corresponding ground truth segmentation masks is notably limited in deep learning terms ($< 10,000$). If animal meat images can improve model performance further, then this may be a way to at least partly negate the difficult problem of wound image acquisition from medical settings. We strongly encourage other researchers working in chronic wound deep learning studies, especially those working in localisation and segmentation, to experiment with such images.

This work is motivated by the development of new technologies that will allow for the remote detection and monitoring of chronic wounds in home settings. Patients living in remote locations have been shown to have worse outcomes when compared to those living in urban areas. The development of new remote monitoring solutions using deep learning techniques may provide a solution to help reduce such disparities (\cite{drovandi2021remotely}). Such technologies have the potential to reduce the number of patient hospital visits, reducing nosocomial infections. The viability of deep learning detection systems within medical settings has been demonstrated for chronic wounds (\cite{cassidy2023eval}). However, further clinical evaluations are required in larger studies to confirm model effectiveness across a more diverse range of skin tones. Such studies will be vital to identify where shortfalls exist in current segmentation models. 

Strategic approaches to preprocessing methods when training deep learning models for chronic wounds have been shown to be highly effective, as per recent work completed by \cite{okafor2024preproc}. This work demonstrates the importance of careful targeting of preprocessing methods for different wound types. Our future work will be guided by these methods to attempt to further improve network performance. 

Future work will focus on models that utilise multi-modal data which will include additional clinical information collected from patient records. These data will include details of infection, ischemia, neuropathy, and other clinical measures such as patient age, ethnicity, and blood type. Work is currently underway with our clinical collaborators to collect the required patient data. Prior studies in similar research domains have shown that multi-modality in training workflows can assist in improvements to model accuracy (\cite{jaworek2021sites}). Using patient IDs linked to dataset images will allow us to reduce the number of cases which are currently spread across training and test sets, reducing the potential biases. We will also expand our work to investigate instance segmentation of wound and periwound to determine if features from surrounding wound tissue can help to improve segmentation and classification accuracy.

We note that there are currently no established standards for the accepted levels of accuracy in chronic wound localisation and segmentation. In general, IoU thresholds of 0.50 and 0.75 are most commonly used (\cite{padilla2021metrics}). However, these measures may differ depending on the research domain. The disparities observed in the present study between lab based metrics and qualitative measures highlight this issue further. We propose that future work should investigate the formulation of accuracy and evaluation standards for chronic wounds via an international consortium of clinical and deep learning experts. The clinical labelling of our datasets reveals that labelling amongst clinicians can be highly variable, a problem which occurs frequently in wound image datasets (\cite{howell2021wound}). Establishing internationally agreed standards may help to improve the accuracy of future models. This is especially pertinent at this stage in the evolution of deep learning models trained using chronic wound datasets, whereby the number of publicly available datasets continues to grow. 

Our research group is currently in the process of capturing video recordings of chronic wounds in medical settings, which we intend on using for future studies. Videos of wounds, captured at different angles would allow for the capture of additional spatial data that may be able to improve the accuracy of predictive models and could be especially useful in the automatic assessment of wound healing over time. Short video clips would be straight forward to capture and analyse using the mobile and cloud frameworks developed in our prior wound studies (\cite{cassidy2021cloudbased, cassidy2023eval}). 



\section{Conclusion}
In this work we proposed a novel harmonic densely connected hybrid transformer network architecture utilising multi-colour space tensor merging. 
We conduct the most comprehensive reliability study to date in chronic wound segmentation using 684 cases to obtain inter-rater reliability measures. 
A total of 13 datasets were used to train and test our proposed segmentation model. 
Our proposed model demonstrates significant improvements over the baseline model in terms of lab based metrics ($+0.1274$ for IoU, $+0.1221$ for DSC) and in terms of expert qualitative assessment (up to 20\% when using a 5 star rating method). 
For the first time, we demonstrate the ability of a model trained only on patients with lighter skin tones to segment wounds on patients with darker skin tones in an effort to address the issue of biases inherent in many chronic wound deep learning studies. 
We also demonstrate performance improvements using GAN-generated wound images and an animal meat dataset in the training workflow. 
The aim of our work is to utilise and build upon state-of-the-art advances in the field to address the problem of accurate chronic wound segmentation and to bring these advances closer to the patients who need them most. 

\section*{Acknowledgments}
We would like to thank clinicians at the King Saud University Medical City, Saudi Arabia for granting permission to use the KSUMC chronic wound dataset in our experiments. We would also like to thank clinicians at the following UK NHS hospitals for providing valuable clinical feedback: Lancashire Teaching Hospitals NHS Foundation Trust, UK; United Lincolnshire Hospitals NHS Trust, UK; Jersey General Hospital, Jersey. Raphael Brüngel was partially funded by a PhD grant from the University of Applied Sciences and Arts Dortmund, Dortmund, Germany.

\bibliographystyle{model2-names.bst}\biboptions{authoryear}
\bibliography{Ref}



\end{document}